\pdfoutput=1
\documentclass[usenatbib]{mnras}
\usepackage{graphicx,amssymb,amsmath,rotating,longtable,yhmath,url,layouts,xcolor,txfonts}
\usepackage[normalem]{ulem}
\usepackage{lastpage}
\usepackage{natbib}
\def\aj{AJ}%
\def\apj{ApJ}%
\def\apjl{ApJ}%
\def\apjs{ApJS}%
\def\ao{Appl.~Opt.}%
\def\aap{A\&A}%
\def\mnras{MNRAS}%
\def\prd{Phys.~Rev.~D}%
\def\nat{Nature}%
\defcitealias{Croom2005}{C05}
\defcitealias{daAngela2008}{dA08}
\defcitealias{Ross2009}{R09}
\defcitealias{Shanks2011}{S11}

\title[The 2QDES Pilot]{The 2QDES Pilot : The luminosity \& redshift dependence of quasar clustering}
\author[B. Chehade et al.]{
\parbox[t]{\textwidth}{
Ben Chehade$^{1}$\thanks{E-mail: ben.chehade@durham.ac.uk }, 
T. Shanks$^{1}$,
J. Findlay$^{1,4}$,
N. Metcalfe$^{1}$,
U. Sawangwit$^{1}$,
M. Irwin$^{2}$,
E. Gonz{\'a}lez-Solares$^{2}$,
S. Fine$^{8}$,
M. J. Drinkwater$^{3}$,
S. Croom$^{6,7}$,
R. J. Jurek$^{9}$,
D. Parkinson$^{5}$,
R. Bielby$^{1}$}
\vspace*{6pt}\\
$^{1}$Centre for Extragalactic Astronomy, Department of Physics, Durham University, South Road, Durham DH1 3LE, U.K. \\
$^{2}$Institute of Astronomy, Univ. of Cambridge, Madingley Road, Cambridge, CB3 0HA, UK \\
$^{3}$School of Mathematics and Physics, University of Queensland, Brisbane, QLD 4072, Australia \\
$^{4}$Department of Physics \& Astronomy, University of Wyoming, 1000 E. University, Department 3905, Laramie, WY 82071, USA \\
$^{5}$School of Mathematics and Physics, University of Queensland, Brisbane, Queensland 4072, Australia \\
$^{6}$Sydney Institute for Astronomy (SIfA), School of Physics, University of Sydney, NSW 2006, Australia \\ 
$^{7}$ARC Centre of Excellence for All-sky Astrophysics (CAASTRO) \\
$^{8}$Department of Physics, University of Western Cape, Bellville 7535, Cape Town, South Africa \\
$^{9}$CSIRO Astronomy and Space Science, Australia Telescope National Facility, PO Box 76, Epping, NSW 1710, Australia \\
\vspace*{-0.5cm}}
\def\LaTeX{L\kern-.36em\raise.3ex\hbox{a}\kern-.15em T\kern-.1667em\lower.7ex\hbox{E}\kern-.125emX}
\begin{document}
\urlstyle{same}
\label{firstpage}
\pagerange{\pageref{firstpage}--\pageref{lastpage}}  \pubyear{2016}
\maketitle
\begin{abstract}
We present a new redshift survey, the 2dF Quasar Dark Energy Survey pilot (2QDESp), which consists of ${\approx}10000$ quasars from ${\approx}150$ deg$^2$ of the southern sky, based on VST-ATLAS imaging and 2dF/AAOmega spectroscopy. Combining our optical photometry with the WISE (W1,W2) bands we can select essentially contamination free quasar samples with $0.8{<}z{<}2.5$ and $g{<}20.5$. At fainter magnitudes, optical UVX selection is still required to reach our $g{\approx}22.5$ limit. Using both these techniques we observed quasar redshifts at sky densities up to $90$ deg$^{-2}$. By comparing 2QDESp with other surveys (SDSS, 2QZ and 2SLAQ) we find that quasar clustering is approximately luminosity independent, with results for all four surveys consistent with a correlation scale of $r_{0}{=}6.1{\pm}0.1 \: h^{-1}$Mpc, despite their decade range in luminosity. We find a significant redshift dependence of clustering, particularly when BOSS data with $r_{0}{=}7.3{\pm}0.1 \: h^{-1}$Mpc are included at $z{\approx}2.4$. All quasars remain consistent with having a single host halo mass of ${\approx}2{\pm}1{\times}10^{12} \: h^{-1}M_\odot$. This result implies that either quasars do not radiate at a fixed fraction of the Eddington luminosity or AGN black hole and dark matter halo masses are weakly correlated. No significant evidence is found to support fainter, X-ray selected quasars at low redshift having larger halo masses as predicted by the `hot halo' mode AGN model of \cite{Fanidakis2013}. Finally, although the combined quasar sample reaches an effective volume as large as that of the original SDSS LRG sample, we do not detect the BAO feature in these data.
\end{abstract}
\begin{keywords}
 \LaTeXe\ - catalogues - quasars:general - galaxies:active - galaxies:Seyfert 
\end{keywords}
\section{Introduction}
Quasars are a very luminous subset of the active galactic nuclei (AGN) population. Due to their high intrinsic luminosities they can be exploited in a wide variety of cosmological and astrophysical studies. Quasars possess an ultraviolet excess (UVX) of emission with respect to stars. The UVX property has previously been exploited by large area surveys to preform quasar selection, such as 2QZ \citep{Smith2005}, 2SLAQ \citep{Richards2005} and SDSS \citep{Richards2002}. In addition to their UVX property, quasars possess an excess of emission with respect to stars in the infra-red. This method of selecting quasars is sometimes known as the KX (K-band excess) technique and has also been used to photometrically select quasars \citep[see ][]{Maddox2012}. The aim of the 2QDESp survey is to maximise the measured sky density of quasars between $0.8{<}z{<}2.5$ and hence minimise the correlation function errors for both cosmological and quasar physics studies.

The 2QZ \citep[hereafter \citetalias{Croom2005}]{Croom2005}, 2SLAQ \citep[hereafter \citetalias{daAngela2008}]{daAngela2008} and SDSS \citep[hereafter \citetalias{Ross2009}]{Ross2009} surveys all measured the quasar correlation function in approximately the same redshift interval ($0.8{<}z{<}2.5$). As each of the surveys had different magnitude limits but similar redshift distributions we can compare these directly to one another to measure the luminosity dependence of quasar clustering. \citet[hereafter \citetalias{Shanks2011}]{Shanks2011} discuss the implications of these three surveys measuring a consistent value of $r_{0}$, the clustering scale. Of particular interest to galaxy formation models is the apparent independence (\citetalias[see][and \citealt{Shen2009}]{Shanks2011,daAngela2008}) of quasar clustering with luminosity. \citetalias{Shanks2011} examines the results from optical clustering measurements (\citetalias{Croom2005,daAngela2008,Ross2009}) and attempts to reconcile these results in context of existing models and finds no evidence for strong luminosity dependence of quasar clustering. This is surprising given the measured relation between optical luminosity and black hole mass, M$_{ \mathrm{BH} }$, \citep{Peterson2004} and black hole mass and dark matter halo mass, M$_{\mathrm{Halo}}$, \citep{Ferrarese2002,Fine2006}. 

However, \cite{Fanidakis2013} predict that less luminous quasars inhabit higher mass halos and reported that  X-ray selected quasars inhabit higher mass halos (${\sim}10^{13}$M$_{\odot}$) than found in optical studies (${\sim}10^{12}$M$_{\odot}$). If optically and  X-ray selected quasars sample distinct populations there may be no contradiction to the conclusions of \citetalias{Shanks2011}. We aim to measure ${\sim}80$ quasars deg$^{{-}2}$, comparable to the sky density quasars in deep  X-ray surveys \citep{Allevato2011}. At these space densities the significant overlap between  X-ray and optical quasar samples should result in larger correlation lengths if  X-ray selected quasars inhabit higher mass halos.

Here we describe the first quasar survey (2QDESp) using VLT survey telescope (VST)-ATLAS \citep{Shanks2014} optical photometry with follow-up spectroscopic observations made using 17 nights of AAT time using 2dF and the AAOmega spectrograph. Combining 2QDESp with several previous quasar surveys we measure the luminosity dependence of quasar clustering for the combined sample. In Section \ref{sec:img} we describe the imaging data and quasar selection techniques. In Section \ref{sec:spec} we describe our spectroscopic follow-up of targets and the resulting quasar catalogue. In Section \ref{sec:qsoclustering} we discuss the methods used to measure quasar clustering and the results from the 2QDESp sample before incorporating the other surveys into our analysis. From this combined sample we make measurements of the luminosity and redshift dependence of quasar clustering in Sections \ref{sec:ldep} and \ref{sec:reddep}. We discuss our results and their implications in Section \ref{sec:dis}. We assume $H_{0} = 100 \: h \: \mathrm{km} \: s^{-1}$ Mpc$^{-1}$ and a flat cosmology from \cite{Planck2014} with $\Omega_{M} = 0.307$. $ugriz$ magnitudes are quoted in the AB magnitude system unless stated otherwise, WISE magnitudes are left in their native Vega system.

\section{Imaging}
\label{sec:img}
\subsection{Imaging}
\subsubsection{VST-ATLAS}
\label{sec:vst}
The VLT Survey Telescope (VST) is a 2.6 m wide-field survey telescope with a $1^{\circ} \times 1^{\circ}$ field of view and hosts the OmegaCAM instrument. OmegaCAM \citep{Kuijken2004} is an arrangement of $32$ CCDs with $2k \times 4k$ pixels, resulting in $16k \times 16k$ image with a pixel scale of $0.21 \arcsec$ . The VST-ATLAS is an ongoing photometric survey that will image ${\approx} 4700$ deg$^{2}$ of the southern extragalactic sky with coverage in $ugriz$ bands. The survey takes two sub-exposures (exposure time varies across filters) per 1 degree field with a $25 \times 85$ arcsecond dither in X and Y to ensure coverage across interchip gaps. The sub-exposures are then processed and stacked by the Cambridge Astronomy Survey Unit (CASU). The CASU pipeline outputs catalogues that are cut at approximately $5\sigma$ and provides fixed aperture fluxes and morphological classifications of detected objects. The $u$-band catalogue comprises `forced photometry' at the position of $g$-band detections; no other band is forced. The processing pipeline and resulting data products are described in detail by \cite{Shanks2014}. Bandmerged catalogues were produced using {\tt TOPCAT} and {\tt STILTS} software \citep{TOPCAT2005,STILTS2006}. Unless otherwise stated, for stellar photometry we use a $1 \arcsec$ radius aperture ({\tt aper3} in the CASU nomenclature). ATLAS photometry is calibrated using nightly observations of standard stars. The calibration between nights can vary by $\pm 0.05$ mag (see Shanks et al. 2015 for details). We performed a further calibration on the fields we observed prior to target selection to ensure agreement between VST-ATLAS fields and the SDSS stellar locus, as described in Section \ref{sec:xdqso}.
With the VST-ATLAS survey under halfway complete during our spectroscopy, the selection of 2dF pointing positions was governed by the progress of ATLAS. The fields are not generally distributed over a spatially contiguous region, although their seeing and magnitude limits are representative of the survey as a whole. 
The morphological star-galaxy classification we use is that supplied as default in the CASU catalogues. This classification is discussed in detail \citep[by][]{Gonzalez2008}. We test the morphological completeness for different colour-colour selections in Section \ref{sec:photoincompleteness}.  
\subsubsection{WISE}
The NASA satellite Wide-field Infrared Survey Explorer (WISE) \citep{Wright2010}, mapped the entire night sky in four passbands between $3.4 - 22 \mathrm{\mu m}$. The survey depth varies over the sky but approximate $5\sigma$ limits for point sources are $W1=16.83$ and $W2=15.60$ mag. in the Vega system. The W1 and W2 bands have point spread functions (PSFs) of $6.1 \arcsec$ and $6.4 \arcsec$ respectively compared with ${\approx} 1 \arcsec$ in the VST-ATLAS bands. A comparison\footnote{http://wise2.ipac.caltech.edu/docs/release/allsky/expsup/sec2\_2.html} between WISE catalogue positions and the USNO CCD Astrograph Catalog (UCAC3) catalogue shows that even at the faintest limits of W1 there is $<0.5 \arcsec$ rms offset between the two catalogues. We matched ATLAS optical photometry to the publicly available WISE All-Sky Source Catalogue using a $1 \arcsec$ matching radius. Given the size of the WISE PSF we examine the possibility of WISE-ATLAS mismatching. For sky density of WISE sources at $|b| > 30^{\circ}$ we calculate that $1$ in $25$ quasars identified in WISE will have a blended WISE source within $3 \arcsec$. Compared to this value, the contribution from quasar-quasar pairs will be smaller. Given the other advantages of using WISE selection, we view this effect as essentially negligible.
\begin{figure*}
\hbox{\hspace{-3cm}}
\makebox[\textwidth][c]{\includegraphics[width=\textwidth]{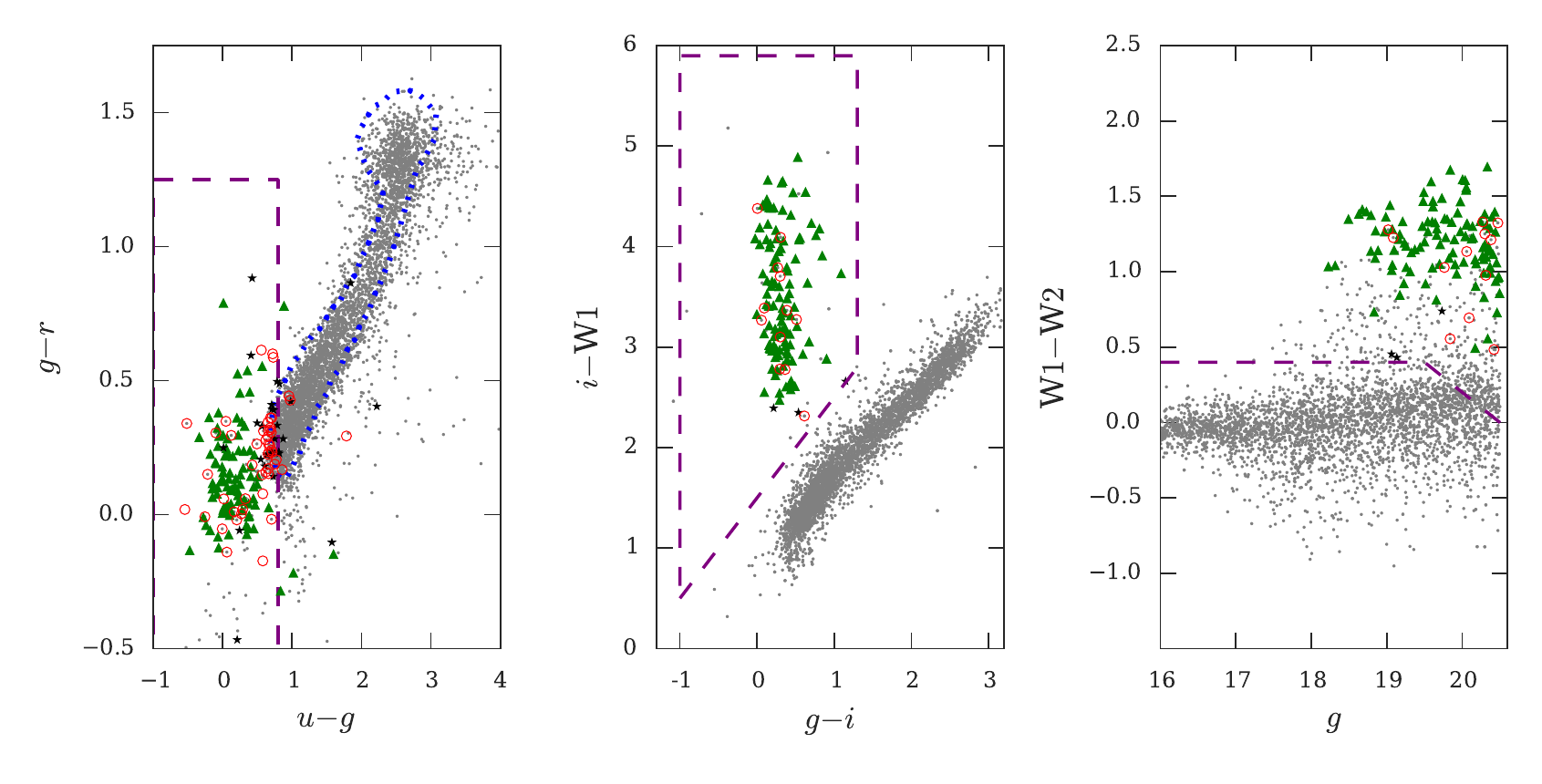}}
\caption{In the left panel we show the $ugr$ colour space of the field centred at $23^h16^m-26^d01^m$. We show all objects identified in the g-band as point-sources (between $16\leq g\leq 20.5$) as grey dots. We show the SDSS Stripe 82 stellar locus (dotted blue line) and our $ugr$ colour cuts (purple dashed lines) from Section \ref{sec:ugri}. Spectroscopically confirmed quasars within our target redshift range ($0.8{<}z{<}2.5$) are shown as green triangles and confirmed stars are shown as black five-point stars. Sources without a positive identification are outlined with a red circle. In the middle panel we show the same objects in the $gi$W$1$ colour space and in the right panel we show the $g$W$1$W$2$ colour space.}
\label{fig:wise_colour_selection}
\label{fig:ugr_giw}
\label{fig:atlastosloan}
\end{figure*}
\section{Other quasar redshift surveys}
\label{sec:qsosurveys}
Here we introduce three additional quasar surveys that were used to measure the clustering of optically selected quasars. To aid comparison between these surveys and our own we summarise the method of quasar selection for each survey, the measured space density, area and size. In Section \ref{sec:qsoclustering} we remeasure the correlation function for these surveys, verifying our measurement against previously published values (see Table \ref{tab:other_surveys}). We then combine these survey with the 2QDESp sample to better constrain the autocorrelation function.
\subsection{2QZ}
\label{sec:2qz}
The 2QZ survey \citep{Croom2004} covers approximately $750$ deg$^{2}$ of the sky in two contiguous areas of equal size. The quasar sample consists of over $22 \: 500$ spectroscopically confirmed sources at redshifts less than $3.5$ and apparent magnitudes $18.25{<}b_{J}{<}20.85$. Quasars are selected based on their broadband optical colours from automated plate measurement (APM) scans from United Kingdom Schmidt Telescope (UKST) photographic plates. Colour selection is performed using $u-b_{J}$ vs. $b_{J}-r$. The measured quasar density is ${\approx} 30$ quasars deg$^{-2}$. 
\subsection{SDSS DR5}
\label{sec:sdss}
The SDSS DR5 uniform sample \citep{Schneider2007} contains $30 \: 000$ spectroscopically confirmed quasars between redshifts $0.3{\leq}z{\leq}2.2$ and an apparent magnitude limit of $i_{SDSS} \leq 19.1$ over ${\approx}4000$ deg$^{2}$. This sample was selected using single epoch photometry from the SDSS using the algorithm given in \cite{Richards2002}. The sample is described in detail by \citetalias{Ross2009} and has a measured quasar density of ${\approx} 8$ quasars deg$^{-2}$. 
\subsection{2SLAQ}
\label{sec:2slaq}
The 2SLAQ survey \citep{Croom2009c} overlaps two subregions within the 2QZ survey area, with an average quasar density ${\approx} 45$ quasars deg$^{-2}$ and redshifts of $z \lesssim 3$. The 2SLAQ survey is based on SDSS photometry and measures redshifts for quasars of apparent magnitudes $20.5{<}g_{SDSS}{<}21.85$. This sample was designed to be used in conjunction with the observations from the 2QZ survey, \citepalias[see ][]{daAngela2008}.

\section{2QDESp quasar selections}
\label{sec:quasarselection}
\subsubsection{Quasar density $g \leq 22.5$}
Previously, 2QZ measured a completeness corrected sky density of $30$ quasars deg$^{-2}$ at $b_{J}{<}20.85$. 2SLAQ reached a nominal density of $45$ deg$^{-2}$ at $20.5{<}g_{\mathrm{SDSS}}{<}21.85$. \citetalias{daAngela2008} combined the 2QZ and 2SLAQ samples to produce a higher density sample of ${\approx} 80$deg$^{-2}$. However, the high incompleteness of 2SLAQ meant this high density was only achieved after the application of completeness corrections. In this survey we aim to measure $80{-}100$ quasars deg$^{-2}$ in the redshift range $0.8{<}z{<}2.5$ in ${\sim}1$ hour 2dF exposures; we demonstrate the feasibility of our aims below.

\subsubsection{Quasar Luminosity Function}
The first concern of the 2QDES pilot is whether or not the luminosity function of quasars predicted $80+$ quasars deg$^{-2}$ within our targeted redshift ($0.8{<}z{<}2.5$) and magnitude ($16{<}g{<}22.5$) range. A small number of quasar redshift surveys have explored this redshift range to fainter limits than 2SLAQ although always in relatively small areas. 
\cite{Fine2012} made a survey based on Pan-STARRS Medium Deep Survey imaging. As well as using colour selection, \cite{Fine2012} also used variability from many epochs of imaging to select their quasar candidates. To a magnitude limit of $g=22.5$ their measured quasar density was $88 {\pm} 6$ deg$^{-2}$ ($0.8{<}z{<}2.5$).

In SDSS Stripe 82 \cite{Palanque-Delabrouille2012} covered ${\approx}15$ deg$^{2}$ and measured a completeness corrected quasar density of $99{\pm}4$ quasars deg$^{-2}$. This was to the same depth as 2QDESp ($g \leq 22.5$) and in a narrower redshift range ($1 \leq z \leq 2.2$). However, both these authors relied on multi-epoch photometry reaching $50\%$ completeness for point sources at $g=24.6$ (c.f. $g{\sim}23$ for VST-ATLAS). 

Finally, spectroscopic follow-up of  X-ray sources in the XMM-COSMOS field \citep{Brusa2010} has measured a quasar density of $110$ quasars deg$^{-2}$ within our redshift interval to a depth of $g{<}22.5$ ($i{\la}22.2$).

Thus from these comparisons to other surveys we can be confident that there exist a sufficiently high space density of quasars within the $g{\leq}22.5$ limit of the survey. However, these complete samples are often selected from deeper imaging with the added benefit of selecting quasars from their variability. 

\subsubsection{Photometric incompleteness from VST-ATLAS}
\label{sec:photoincompleteness}
The second question we address is whether the VST-ATLAS catalogues are of sufficient depth to select $80{-}100$ quasars deg$^{-2}$. As an approximate test of our photometric completeness we rely on the $u$ and $g$-bands where quasars have the colour $u{-}g{<}0.5$. In the $g$-band the limit $g{<}22.5$ which is ${\sim}0.7$ mag brighter than the median $5\sigma$ depth of VST-ATLAS as such we assume we are always complete in this band. The $5\sigma$ limits are based on sky noise but as the $u$-band is forced this limit may not provide a good estimate of the image depth. We match to the deeper KIDS survey \citep{deJong2013} in an area of VST-ATLAS with $u_{5\sigma}=21.7$ \citep[$90\%$ of tiles have fainter limits, see][]{Shanks2014}. We find that the use of forced photometry in the $u$ results in $50\%$ completeness (c.f. KIDS) at $u{=}22$ which is $0.3$ mag deeper than the $5\sigma$ limit. Applying the limits $g{<}22.5$ and $u{<}22$ (with $0.8{<}z{<}2.5$) to the \cite{Fine2012} data we recover $80{\pm}6$ quasars deg$^{-2}$. Assuming median depth limits ($u{=21.99}$) gives $87{\pm}6$ quasars deg$^{-2}$.

The number counts of the Fine et al. suggest the sample is complete to $g{\approx}21.9$. This incompleteness at fainter magnitudes will lower the estimated return of quasars in the VST-ATLAS data. We note that the more complete data of \cite{Palanque-Delabrouille2012} returns ${\sim}10\%$ more quasars than \cite{Fine2012}. 

We have taken a conservative approximation of our photometry and a conservative estimate of the true quasar density and found the VST-ATLAS photometry is sufficiently deep to return our minimum target density ($80$ quasars deg$^{-2}$). By assuming more representative photometry and comparing to a more compete quasar sample we expect these estimated densities to increase. We finally note that in the 2QDESp survey (see Sections \ref{sec:qsocat} and Table \ref{tab:summary}) that $90\%$ of 2dF pointings have a $u_{5\sigma}{>}21.85$ i.e. the $u$ imaging is slightly better than found in VST-ATLAS as a whole.

\subsubsection{ugri selection}
\label{sec:ugri}
The UVX property of quasars was successfully used by both 2QZ and SDSS to select quasars in our target redshift range ($0.8{<}z{<}2.5$). As our photometric bands are the same as those used by SDSS, we can base our colour selection on work from the SDSS collaboration. We used known 2QZ quasars within the ATLAS footprint to determine the colour cuts suitable for use with VST-ATLAS aperture photometry. 

For reference, we show the location of our cuts in $ugr$ colour space in the left panel of Figure \ref{fig:atlastosloan}. Our selection criteria were as follows;
\begin{itemize}
\item $ -1 \leq (u - g) \leq +0.8$
\item $-1.25 \leq (g - r) \leq 1.25$
\item $(r - i) \geq 0.38 - (g - r)$ 
\end{itemize}
We applied this colour selection in a 2dF field with typical VST-ATLAS depth and seeing and found ${\approx} 600$ candidates deg$^{-2}$, where we considered only point sources (in the $g$-band) and targets between $16{<}g{<}22.5$. These cuts selected a large area in colour space (minimising the effect of colour incompleteness) and therefore resulted in a high sky density of targets but with significant stellar contamination. We relied on the combination of these cuts and the {\tt XDQSO} algorithm (see Section \ref{sec:xdqso}) to minimise this stellar contamination, particularly for fainter targets $21.5{<}g{<}22.5$.

Due to the proximity of the quasar locus to main sequence (MS) stars, photometric errors are a concern for optical quasar selection. Galaxies may be incorrectly identified as point-sources from their morphology but galaxy colours are sufficiently non-quasar like that galaxy contamination is not considered to be problematic. Morphological incompleteness may be introduced, however, by identifying point-sources as extended sources and therefore not selecting them as quasar candidates. To mitigate this effect we rely on the deeper VST-ATLAS bands to perform our morphological cuts (the $g$ and $r$-bands). We relied on two bands to account for the possibility of poor image quality affecting the morphological classification in a single band. 

\subsubsection{Optical and mid-IR selection}
\label{sec:lx}
By combining the mid-infrared photometry from WISE with the optical bands from VST-ATLAS we achieve a larger separation between the stellar locus and our target quasars than is possible using optical colours alone (see Figure \ref{fig:wise_colour_selection}). Unlike the $ugri$ colours this selection relies on the infrared excess of emission to differentiate between stars and quasars. 

Quasars in our target redshift range have a mean $g-$W$1 = 4$ with a large dispersion ${\approx} \pm 1$. The $5\sigma$ limits are $g {\approx} 23.25$ and W$1{\approx} 16.83$. As such the depth of our mid-IR selection is limited by the depth of the WISE photometry.

In the centre panel of Figure \ref{fig:wise_colour_selection} we show the $g-i$ colour plotted against the $i-$W$1$ colour. The stellar locus is clearly identifiable. The right hand panel shows the mid-IR colour W$1{-}$W$2$ as a function of g band magnitude. The latter colour selection helped to remove any remaining stellar contamination that was left by the $g{-}i:i{-}$W$1$ colour cut. 

The colour cuts we applied are given here;
\begin{itemize}
\item $ (i - $W$1) \geq (g - i) + 1.5 $
\item $-1 \leq (g - i) \leq 1.3 $
\item $ (i - $W$1) \leq 8$
\item $($W$1 - $W$2) > 0.4 \:\: \& \:\: g < 19.5$
\item $($W$1 - $W$2) > -0.4g + 8.2 \:\: \& \:\: g > 19.5$
\end{itemize}
Within a typical 2dF pointing, this target selection returns ${\approx}100$ candidates deg$^{-2}$. This algorithm therefore supplies optimal target density to observe all candidates on the 2dF in a single exposure. However, to meet our target density we required this algorithm to be both highly complete and free from contamination. 
We used the $gi$W$1$ colours to test our morphological classification of sources from their g-band imaging, by separating the galaxy and stellar loci in colour space. We determined that of the stars identified by their colour, $91.5 \pm 0.5 \%$ were identified as point sources by the $g-$band morphological classification. We tested the morphological classification in the r-band with the $rz$W$1$ colours and found a similar value. 

\subsubsection{XDQSO Algorithm}
\label{sec:xdqso}
Automated quasar selection algorithms typically compare broadband colours to model quasar colours (or some library of previously observed quasars). As the VST-ATLAS survey has the same filter set as the SDSS survey, there exists a legacy of quasar selection code \cite[such as][]{Richards2004,Kirkpatrick2011,Bovy2011}. \cite{Bovy2011} demonstrated the success of the {\tt XDQSO}\footnote{v0\_6} algorithm and we applied this algorithm throughout our observing program.
The {\tt XDQSO} algorithm takes as input the colours of a source and compares this to empirical observations of quasars and stars. The code outputs a relative likelihood (${\tt PQSO} \: \epsilon [0,1]$) that an object is either a star or a quasar. The {\tt XDQSO} algorithm uses SDSS as its training data and so we must consider both colour terms between SDSS and VST-ATLAS filters and differences in photometric zeropoints. If these differences between SDSS and VST-ATLAS are small then we shall be able to implement the {\tt XDQSO} algorithm without modification. At the time of our spectroscopic programme, VST-ATLAS photometry was supplied in the Vega system. To convert to the SDSS system we adjusted the zeropoints of the individual VST-ATLAS tiles to get good agreement with the SDSS Stripe 82 coadd photometry for stars. In the left panel of Figure \ref{fig:atlastosloan} we show the outline of the stellar locus from Stripe 82; the VST-ATLAS photometry is seen to be in good agreement with this deeper photometry. We refer the reader to \cite{Shanks2014} where the SDSS-VST colour terms are shown to be small.

The output of this selection algorithm is continuous and assigns candidates with a relative quasar likelihood. As such, we are not limited by a lack of candidates but by the availability of instrument fibres. Whilst the precise sky density of {\tt XDQSO} candidates varies with image quality (and hence sky position), selecting candidates ranked according their {\tt PQSO} value limits us to observing candidates with {\tt PQSO} $\gtrsim 0.7$.

\section{Spectroscopic Observations}
\label{sec:spec}
\subsection{2dF \& AAOMEGA}
Spectroscopic observations were made with the 2dF-AAOmega instrument on the AAT. The 2dF is a fibre positioning system for the AAOmega multi-object spectrograph which is capable of simultaneously observing $392$ objects over ${\approx}3.14$ deg$^{2}$ field of view. Fibres are positioned by a robotic arm and are fed to the  spectrograph. The 2dF also implements a tumbling system that allows for a second plate to be configured whilst the first is being observed.
AAOmega is a dual beam spectrograph that utilises a red/blue dichroic beam splitter, splitting the light at $5700\makebox{\AA}$. The observations were made using the $580V$ and $385R$ gratings for the blue and red arms of the spectrograph respectively. The gratings have resolving power of $R{=}1300$ and central wavelengths of $4800\makebox{\AA}$ and $7250 \makebox{\AA}$ for the blue and red arms. The useful wavelength range in our configuration is $ 3700 \makebox{\AA} $ to $ 8800 \makebox{\AA} $.

We made no nightly observations of standard stars so our spectra do not have an accurate absolute calibration. The {\tt \sc 2DFDR}\footnote{http://www.aao.gov.au/science/software/2dfdr} data reduction pipeline combines the spectra from the red and blue arms of the spectrograph. To achieve this, the spectra are calibrated to a common scale with an arbitrary normalisation due to unknown aperture losses, via a transfer function derived from previous observations of the standard star EG 21.

Of the $392$ 2dF fibres (not including $8$ for guide stars) $20$ fibres for sky subtraction. The remaining ${\approx} 372$ fibres were used for science targets. The fibre allocation software {\tt \sc configure}\footnote{http://www.aao.gov.au/science/software/configure} (v7.17) allows input targets to have priorities associated with them. The observing priorities allow the software to make a decision about the importance of placing fibres. This allowed us to prioritise our spectroscopic targets according to their likelihood of being a quasar. This prioritisation was one of the requirements of the target selection process.

Exposure times varied between $0.7-2$ hours to account for observing conditions. All our data was reduced using the {\tt 2DFDR } pipeline (v5.35) using default parameters. We measured quasar redshifts of the spectra with the {\tt \sc runz } programme \citep{Saunders2004}.
\subsection{Resulting QSO catalogue}
\label{sec:qsocat}
\begin{figure}
\includegraphics[width=\linewidth]{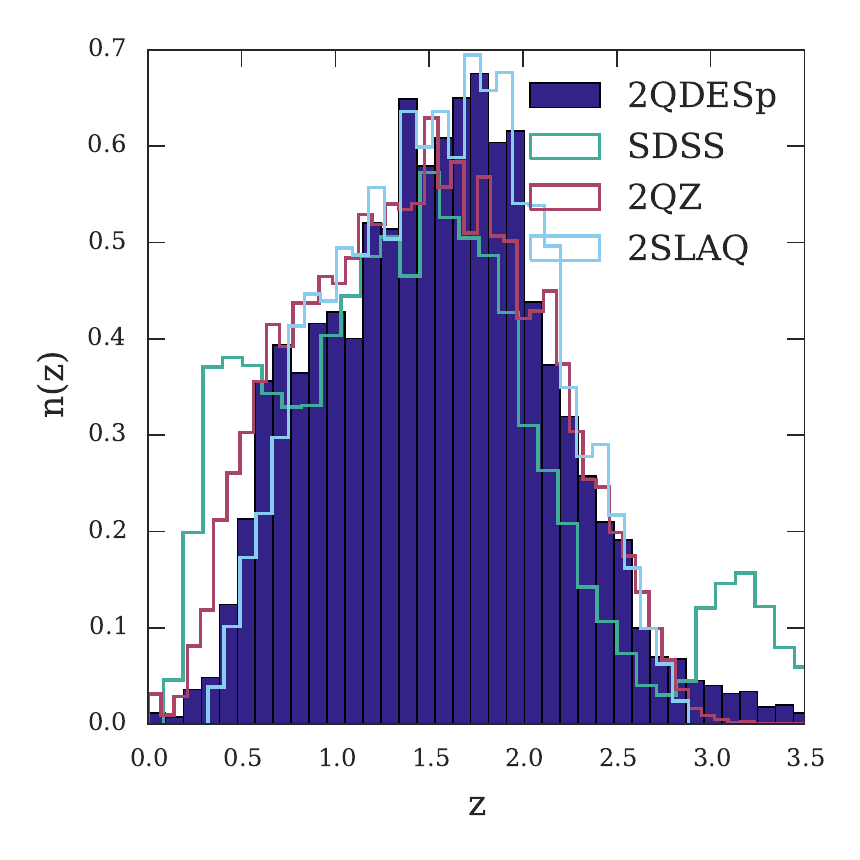}
\caption{We show the redshift distribution of the 2QDESp spectroscopic quasar sample as the shaded region. For comparison we include the redshift distributions for the SDSS DR5, 2SLAQ and 2QZ samples.}
\label{fig:nz}
\end{figure}
We developed a combination of the three techniques described in Section \ref{sec:quasarselection} to optimise our quasar selection over the duration of the pilot survey. We divided the selection into two major implementations. The first (chronological) selection relied solely on the optical photometry from VST-ATLAS in the form of $ugri$ and {\tt XDQSO}. Later implementation of the selection algorithm applied these techniques in conjunction with optical-IR colour cuts. We label these algorithms in Table \ref{tab:summary} as ``$ugri${\tt XDQSO}'' and ``$ugri${\tt XDQSO}W1W2'' respectively. 

The ``$ugri${\tt XDQSO}'' target selection was based on $ugri$ colours with the {\tt XDQSO} algorithm used to rank those targets. The ``$ugri${\tt XDQSO}W$1$W$2$'' selection algorithm gave candidates meeting the optical$+$IR conditions from Section \ref{sec:lx} the highest priority, with remaining candidates prioritised based primarily on their {\tt PQSO}. 

In Table \ref{tab:summary} we present the results of our spectroscopic observations. We list the field locations and the dates of our observations, the number of quasars identified in a given pointing, exposure times, mean spectral signal-to-noise and a guide to the target selection used. 

As variation in spectral signal-to-noise affects our measured quasar density, we look to parametrise our different fields in a meaningful manner so that we can compare the effectiveness of our selection algorithms. To account for varying spectroscopic incompleteness between different observations we compare the number of bright $g{\leq}20.5$ quasars between our fields (see column ``$N_{QSO}{\leq}20.5$'' in Table \ref{tab:summary}). At these brighter magnitudes we are approximately spectroscopically complete for quasars. We show in Table \ref{tab:summary} the faint limit in the $g$-band that contains $90\%$ of our spectroscopically confirmed quasar sample and see that we are suitably bright to be photometrically complete in the $ugriz$ bands.

Having accounted for spectroscopic incompleteness we expect variation in our measured quasar density to be primarily determined by our selection algorithm and the number of background stars (see Section \ref{sec:stellar_density}). Whilst the VST-ATLAS is limited to high galactic latitudes we note that the number of stars in a 2dF pointing can vary significantly (see Table \ref{tab:summary}). We include in Table \ref{tab:summary} the number of point sources of magnitude $g{\leq}21.5$ under the heading $N_{stars}$ and see this density vary by up to a factor of three. This variation is primarily determined by galactic latitude (c.f. ${\approx}5\%$ from zeropoint errors).

Our spectroscopic programme was awarded $17$ nights of observing time to develop a QSO selection algorithm as preparation for a larger programme. We obtained redshifts for ${\approx}10 \: 000$ quasars with apparent magnitudes $g{\leq}22.5$ and $<z>=1.55$ ($80\%$ of the sample lies within $0.8{<}z{<}2.5$). We present the redshift distribution of our quasars in Figure \ref{fig:nz} and include the redshift distributions for 2QZ, 2SLAQ and SDSS for comparison. We see that the SDSS $n(z)$ has a second peak at $z{\sim}3.1$ which is due to a secondary colour selection designed to identify quasars at this redshift. When comparing between surveys we limit to $0.3{<}z{<}2.9$ and so ensure good agreement between the redshift distributions.
\subsubsection{Redshift errors}
\label{sec:redshifterror}
Here we consider factors which affect the measurement of quasar redshifts. Poor quality spectra will cause errors in redshift estimation as well as incompleteness due to failure to identify the target as a quasar. Reliance on a small number of quasar emission lines will also cause systematic errors due to mis-identification of emission features, as noted by \cite{Croom2004}. We have a number of repeat observations that we can use to quantify the redshift error. The {\tt \sc runz} code allows for three quality flags {\tt qop = 5, 4 or 3} for reliable redshifts. 

Restricting our analysis to the highest quality spectra (${\tt qop}=5$) we find a redshift error of $\sigma(z)/z=0.002$, comparing repeat observations as in \cite{Croom2004}. This corresponds to ${\sim}600 \: \mathrm{km} \: s^{-1}$ or ${\sim}2 h^{-1}$ Mpc (comoving) at our mean redshift. We next compare the highest quality to the intermediate quality ({\tt qop = 4}) spectra. We take any difference in redshift greater than $\Delta z = 0.01$ as a redshift failure. Intermediate quality spectra have a redshift failure rate of $6\pm2\%$ and an error of $\sigma(z)/z=0.002$, excluding $\Delta z{>}0.01$. Similarly we find $\sigma(z)/z=0.002$ for our lowest quality spectra ({\tt qop = 3}) but this time with a failure rate of $16{\pm}12\%$.

We quantify the rate of redshift failure due to line mis-identification. Having examined quasars with repeat observations we find that redshift failures occur for $9{\pm}1\%$ of quasars, over all redshifts, magnitudes and spectral quality. 
\subsection{Effectiveness of Quasar selection methods}
\label{sec:effective}
We introduced the optical-IR colour cuts in Section \ref{sec:lx}; we review their effectiveness here and compare to the {\tt XDQSO} technique which relies on UVX techniques to identify quasars. To compare these selection techniques we examine one of the most complete fields ($23^{h}16^{m}-26^{d}01^{m}$) where we find over $80$ quasars deg$^{-2}$ between $0.8{<}z{<}2.5$ and $16{<}g{<}22.5$.

In Figure \ref{fig:ugr_giw} we show the distribution of our quasar sample in the $ugr$, $gi$W$1$ and $g$W$1$W$2$ colour spaces. We include only point sources with $g < 20.5$. As noted in section \ref{sec:lx} the left and middle panels of Figure \ref{fig:ugr_giw} show a difference in distance between the quasars and the stellar locus. At fainter magnitudes ($g>20.5$) the photometric scatter will become larger and so that the effective separation between the stars (mainly type A and F) and quasars will be reduced. The increased distance from the stellar locus seen in $gi$W$1$ (compared to the distance in $ugr$ colour space) suggests that this selection may suffer from less stellar contamination than using $ugr$ photometry and that any stellar contamination might come from different spectral types of stars. 

We examine the apparent purity of the $gi$W$1$W$2$ colour selection by comparing its effectiveness against the {\tt XDQSO} algorithm. We are limited in the $giW1W2$ selection by the depth of the WISE photometry and so must take this into account when comparing to the {\tt XDQSO} algorithm. We limit our comparison to $g{<}20.5$ and treat photometric and spectroscopic incompleteness for quasars as negligible. 
In Table \ref{tab:highly_completefield} we take all $giW1W2$ quasar candidates with $g<20.5$ and find that $3\%$ of these sources are stars, based on spectroscopic observations. If we assume all of the non-identified sources are stars, our stellar contamination rises to $14\%$. We test {\tt XDQSO} by taking the same target density as identified by $giW1W2$ and find contamination rates of $17-42\%$, again depending on the nature of the non-ids.

Within our target redshift range we expect to find ${\simeq}75$ quasars per 2dF pointing at this ($g{<}20.5$) magnitude limit. In Table \ref{tab:highly_completefield} we show the number of quasars identified by both algorithms as well as showing (in brackets) the number of quasars common to both. In the brighter regime ($g{<}20.5$), we find that both algorithms identify at least $74$ quasars within our target redshift interval and so both are consistent with being complete. However, we also note a further $9$ quasars from the $giW1W2$ selection which corresponds to a $12\%$ increase against the performance of {\tt XDQSO}. 

The single quasar ($g<=20.5$ \& $0.8{<}z{<}2.5$) ``missed'' by $giW1W2$ is not detected in the All-Sky release and so was not missed due to incompleteness introduced as a result of our chosen colour selection. However, subsequent to our observations, an improved analysis of the WISE data (the ALLWISE data release) results in a detection for this target ($W1=17.07, S/N=8.5$ and would be selected by our algorithm). This missing target suggests that our WISE photometry has an incompleteness for quasars within our target redshift range of ${\sim}1\%$.

Some quasars are only identified by $gi$W$1$W$2$ but not by {\tt XDQSO}. Many of these would be found by our $ugr$ colour selection, or a simpler colour-magnitude cut. The mean ``probability'' of these targets is ${\tt PQSO}=0.3$ and so would not be observable without a much higher fibre density. The {\tt XDQSO} algorithm provides a ``{\tt PQSO\_ALLZ}'' parameter, that gives the ``likelihood'' of a target being a quasar at any redshift. For these ``missed'' quasars the mean ``likelihood'' is {\tt PQSO\_ALLZ}$=0.8$. The low {\tt PQSO} of these quasars is apparently caused by {\tt XDQSO} attempting to estimate the redshift of the quasar candidates. The two red ($u{-}g{>}0.7$) quasars detected by WISE are given low ratings by {\tt XDQSO} (${\tt PQSO\_ALLZ}=0.81,0.01$ and ${\tt PQSO}=0.26,0.01$) and are therefore too lowly ranked by {\tt XDQSO} to be observed.  

In the fainter regime, {\tt XDQSO} identifies a single quasar redder than $u{-}g{=}0.7$ of the $74$ quasars within this redshift range. The $gi$W$1$W$2$ selection recovers $9$ (out of $78$) with red optical colours. Combining the two selections we find that $9\%$ of our sample in the fainter regime is ``reddened'' beyond the approximate limits of our $ugri$ colour selection.

We widen the redshift interval from $0.8{<}z{<}2.5$ to $0.3{<}z{<}3.5$, in the expectation that errors in redshift estimation performed by {\tt XDQSO} will result in a high number of quasars outside the $0.8{<}z{<}2.5$ interval. We find that both algorithms select a significant number of quasars outside our targeted redshift range. For astrophysical studies of the quasar population this may not be an issue. However, to make the highest precision measurements of clustering, surveys require the highest density of quasars within as narrow a redshift interval as possible. Targeting quasars outside our preferred redshift interval lowers the efficiency of the survey.  
\begin{table}
\resizebox{\columnwidth}{!}{%
\begin{tabular}{l|llll}
\hline
\hline
Selection & Spectroscopic I.D.& & & \\
\hline
		        &QSO       & QSO       & Stars& Non-id  \\
                    &\footnotesize{$(0.8{<}z{<}2.5)$} & \footnotesize{$(0.3{<}z{<}3.5)$}	&      &         \\
$giW1W2$\textdagger& 84 (74)       & 106 (85)	        & 3    & 12      \\
${\tt XDQSO}$\textdagger&75	        & 86                    & 15   & 21             \\
\hline
$giW1W2$\textdaggerdbl&78   (39) & 84 (40)          & 4    & 86          \\
${\tt XDQSO}$\textdaggerdbl&74            & 77            & 4    & 93           \\
\hline\\

\end{tabular}%
}
\caption{Here we show the relative performance of the XDQSO against a $gi$W$1$W$2$ colour cut in a single 2dF with our highest completeness. We divide our comparison of the two algorithms into brighter objects $16{<}g{<}20.5$ (denoted by \protect{\textdagger}) and fainter objects $20.5{<}g{<}22.5$ (denoted by \protect{\textdaggerdbl}). Numbers are deg$^{-2}$ and bracketed numbers show the number of quasars common to both selections.}
\label{tab:highly_completefield}
\end{table}

We now examine spectroscopically confirmed quasars that are ranked as likely stars by the {\tt XDQSO} algorithm. Given the continuous nature of the likelihood we make a cut in the output likelihood. We cut at ${\tt PQSO} < 0.4$ as the target density above this value is ${\approx}250$ degree$^{-2}$, well above what is observable in a single epoch of spectroscopy with 2dF. After the $gi$W$1$W$2$ selection, we find ${\simeq}10$ quasars deg$^{-2}$ (across all magnitudes), within our targeted redshift range that lie within this low {\tt PQSO} region. 

The mid-IR excess demonstrated by quasars places them in a region of colour space with a lower contamination rate than we see from {\tt XDQSO}. If this contamination rate were to continue to fainter limits, then a mid-IR selection alone may be sufficient to achieve the target quasar density. With the current limits from the mid-IR photometry, which introduce photometric incompleteness, we must supplement the mid-IR selection with {\tt XDQSO} to achieve higher quasar densities. In our sample field {\tt XDQSO} recovers a quasar density of 54 quasars deg$^{-2}$ ($0.8{<}z{<}2.5$, $g{<}22.5$) compared to 74 quasars deg$^{-2}$ from combining WISE, VST\-ATLAS and {\tt XDQSO}.
\subsection{The nature of mid-IR non-ids}
Here we examine the contaminants within the $gi$W$1$W$2$ colour space and attempt to discern the nature of the unidentified targets. We look both at the confirmed contaminants from a highly complete field and at the contaminants from the whole survey. Within the highly complete region (from Section \ref{sec:effective}) we have three spectroscopically confirmed stars with $g{<}20.5$. These stars are identified as A and K-type stars from their spectroscopy and have been scattered up from the stellar locus. They are anomalously red in the $i-$W$1$ colour and hence included within our colour selection.
Over the course of the survey we identified a number of white dwarfs (WDs) and M-type stars within our $gi$W$1$ colur space. However, neither of these type of stars have broadband colours consistent with being identified by our $gi$W$1$ selection. WDs have colour $i-$W$1{\lesssim}1$ and M stars have $g{-}i{>}1.5$ so neither of these should contaminate the $gi$W$1$ colour space. 

\begin{figure}
\includegraphics[width=\linewidth]{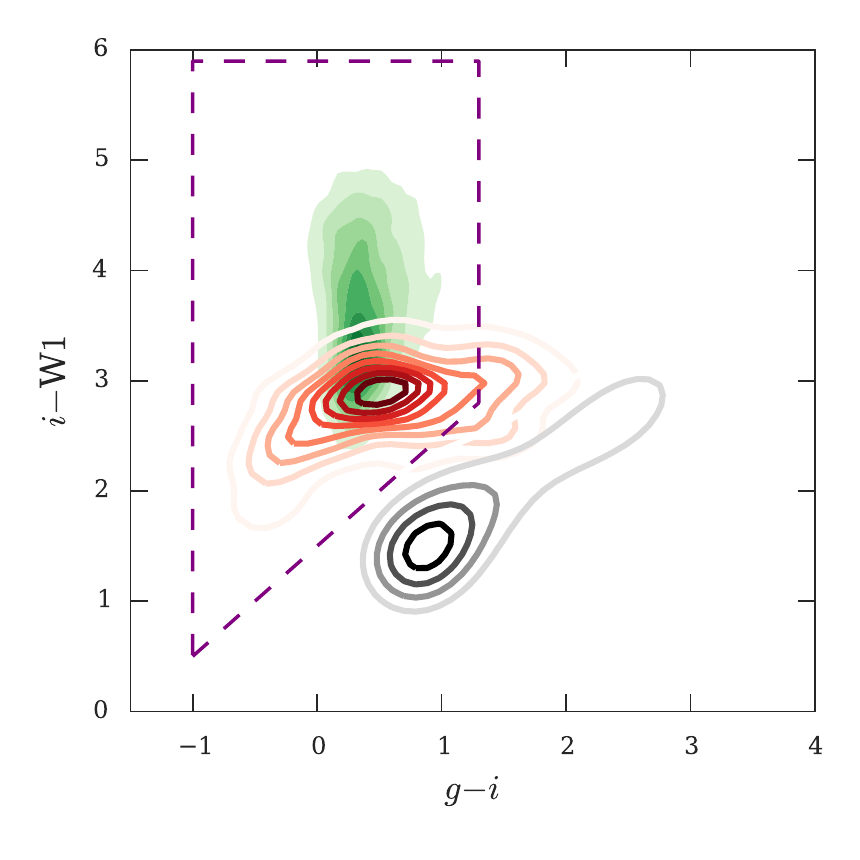}
\caption{We show the distribution of our spectroscopic quasar sample, from the entire survey, in the $gi$W$1$ colour space (green contour). We include morphological point-sources (identified by the g-band with $16{<}g{<}20.5$; shown as grey contour) and our $gi$W$1$ colour cuts (dashed purple lines) from Section \ref{sec:lx} for reference. We show that the WD$+$M binaries from \protect \cite{Debes2011} (red contour) directly overlap with the quasar locus in this colour space acting to reduce the efficiency of this colour selection.}
\label{fig:wd}
\end{figure}

The WIRED survey, \cite{Debes2011}, categorised the infrared emission from UKIDSS Z-band to WISE W4 band of SDSS DR7 WDs. WDs with an infrared excess (mostly due to a contaminating M star) were identified as a potential source of the observed WD contamination of our colour space \citep[see][]{Debes2011}. In Figure \ref{fig:wd} we show that the WD$+$M star locus overlaps with the quasar locus in the $gi$W$1$ colour space. 

\cite{Debes2011} suggested that this may be due to the M-star contributing flux at longer wavelengths than the WD and thereby giving the system quasar colours. These authors found that $28\pm3\%$ WDs have M dwarf companions and approximately a further ${\approx}2{-}10\%$ have either associated dust or a brown dwarf that would give them excess emission in the W$1$ band. Given the similar depths between SDSS and ATLAS we expect a similar rate of contamination as found by \cite{Debes2011}. 

\begin{figure*}
\includegraphics[width=\textwidth]{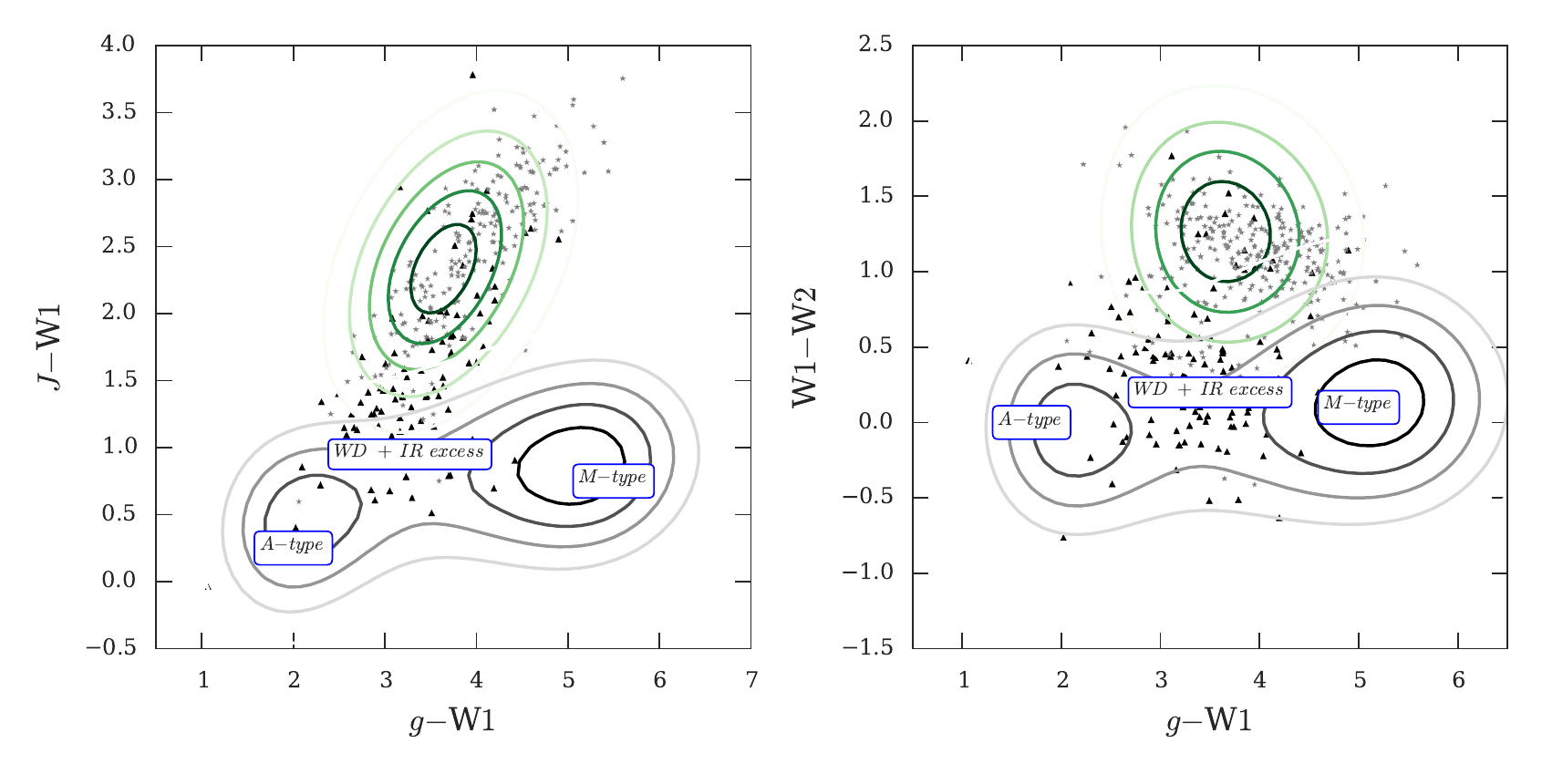}
\caption{In the left panel we show the stellar locus in $gJ$W$1$ colour space (grey contours). We also plot the locus of our quasar sample (green contours). Targets without a spectroscopic id that fulfil the $gi$W$1$W$2$ colour cuts are shown as grey five point stars and spectroscopically confirmed stars are shown as black triangles. We also mark the location of spectral type A and M stars as well as the location of WD$+$IR excess stars from \protect \cite{Debes2011}. In the right panel we follow the same convention for marking the quasar and stellar locus, but instead show these in the $g-$W$1$ vs. W$1-$W$2$ colour space. The majority of non-ids have colours consistent with quasars in $giJ$W$1$W$2$ photometric bands and suggest that the failure to positively identify these targets is due to spectroscopic incompleteness.}
\label{fig:gwj}
\end{figure*}

To better examine the overlap between the quasar locus and the WD$+$M-star locus, we take our entire quasar sample and map its distribution in $gi$W$1$ colour space in Figure \ref{fig:wd}. We show that the WD$+$M binaries directly overlap with the quasar locus in this colour space. Figure \ref{fig:wd} explains the appearance of WDs and M stars in the $gi$W$1$ colour cut. Whilst these systems will account for the occasional appearance of the spectroscopically confirmed contaminants, however, they do not have a sufficiently high sky density to account for the majority of the unidentified sources. 
We attempt to find a colour space that separates the stars and quasars that appear in the $gi$W$1$W$2$ colour selection. If we are able to separate the stars and quasars by colour selection we may be able to infer the nature of the unidentified targets. 
We use all our observed targets to better characterise the contamination. In the right panel of Figure \ref{fig:gwj} we show the $g$W$1$W$2$ plane. We show as reference the position of the stellar locus, the locations of spectral type A and M stars, as well as the WDs with excess IR emission. We also show the quasar locus (derived from our whole sample). Spectroscopic stars and non-ids with $g<20.5$ that obey our $gi$W$1$W$2$ colour selection are also plotted.
\begin{table}
\begin{tabular}{l|lll}
\hline
\hline
 			        &QSO                            & Stars         & Non-id        \\
                    &\footnotesize{$(0.8{<}z{<}2.5)$}   &               &               \\
\hline
$J-W1 > 1.5$     &  $98\%  \: (3583)$              &  $22\% (126)$ & $92\% (1311)$   \\
$J-W1 < 1.5$     &	$ 2\% \: (2)$                  &  $78\% (460)$ & $ 8\%  (111)$ \\
\hline
$J-W1 > 2  $     &   $83\%$            &  $13\%$   & $80\%$   \\
$J-W1 < 2  $     & 	 $17\% $           &  $87\%$   & $20\%$ \\
\end{tabular}
\caption{ We show that the distribution in the $J-W1$ colour for spectroscopic quasars, stars and non-ids. At two different cuts, the distribution of the non-ids more closely follows that for quasars. As such, we infer that the greater part of the non-identified sources are quasars that are not positively identified by our spectroscopic observations.}
\label{tab:w1j}
\end{table}
We see from the right panel of Figure \ref{fig:gwj} that the majority of stars may be removed by a cut in the W$1{-}$W$2$ colour. Due to the close proximity of the stellar locus, a cut in this colour may improve our efficiency but will affect our completeness as well. Where we have J-band coverage from the VHS survey \citep{McMahon2013}, we see from the left panel of Figure \ref{fig:gwj} that the $J{-}$W$1$ colour increases the separation between the stellar locus and the quasar locus. The majority of the non-ids lie within the quasar locus, although some do lie closer to the stellar locus. 
Table \ref{tab:w1j} quantifies that their mean $J{-}$W$1$ colours are consistent with quasar colours. Given that the number counts for non-ids peak a magnitude fainter 
than the peak of the identified targets, this suggests that many of these non-ids are quasars where positive identification has failed due to spectroscopic incompleteness.
\begin{figure}
\includegraphics[width=\columnwidth]{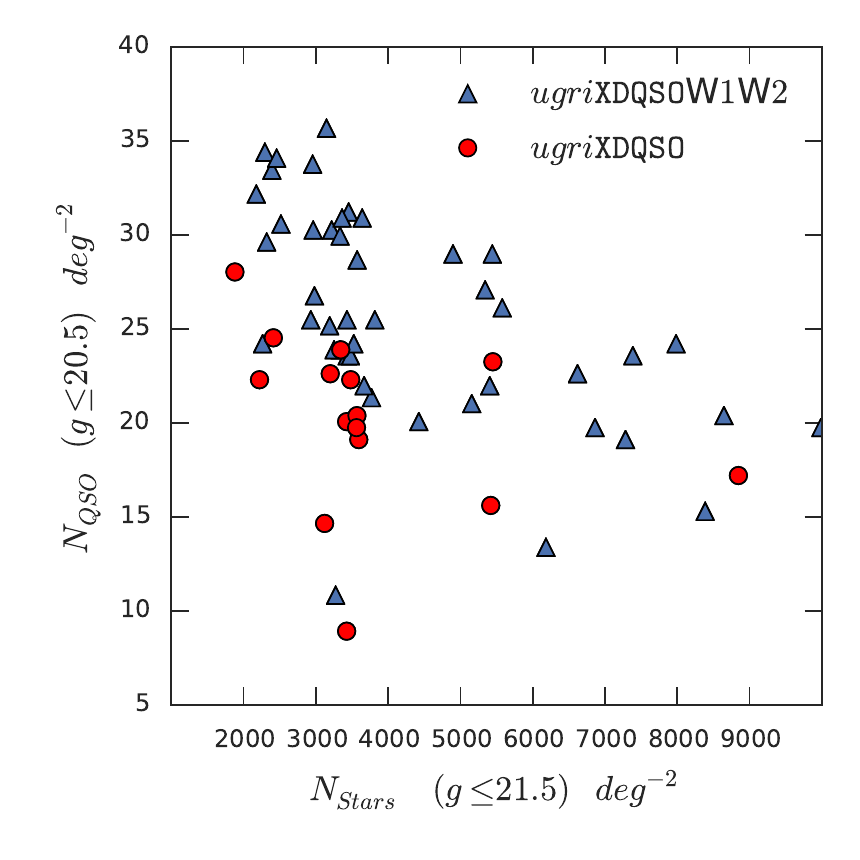}
\caption{We show the number of confirmed quasars deg$^{-2}$ ($g{<}20.5$ ) against the number of stellar sources deg$^{-2}$ (with $g<21.5$ ). We compare the two algorithms; $ugri${\tt XDQSO} (red circles) and $ugri${\tt XDQSO}W$1$W$2$ (blue triangles). By limiting the comparison to brighter quasars we assume the contribution of observational effects is negligible.}
\label{fig:spec_results}
\end{figure}
\subsection{Background stellar density}
\label{sec:stellar_density}
We find that the measured spectroscopic quasar density varies across the sky, independently of the selection algorithm (see Table \ref{tab:summary}). Spectroscopic incompleteness will contribute to this variation. We minimise this by considering only sources with $g{<}20.5$, where we are approximately complete. In Figure \ref{fig:spec_results} we show the variation of confirmed quasars with $g{<}20.5$ as a function of the number of point sources with $g{<}21.5$ per square degree. 

Figure \ref{fig:spec_results} shows that the inverse correlation between stellar density and spectroscopic quasar density is the dominant cause of varying quasar density across the sky. This affect results in as much as a factor of two between confirmed quasars in fields with different background stellar densities. This relation indicates that the efficiency of a wide area quasar survey depends on the background stellar counts in the observed fields.

However, we also note that the $ugri${\tt XDQSO}W$1$W$2$ algorithm fields consistently identify a higher number density of quasars then the $ugri${\tt XDQSO} algorithm. 
\subsection{Conclusions}
We combined mid-IR photometry from WISE with the $ugriz$ photometry from the VST-ATLAS survey to improve the efficiency of our quasar selection. We found that the $giW1$ colour space (see Figure \ref{fig:wise_colour_selection}) is approximately complete to $g{<}20.5$. Fainter than this the colour space becomes photometrically incomplete as quasars become too faint to be detected in the W1 band of WISE. We next attempted to use broadband colours to identify fainter (in the $g$-band) candidates from the $g{-}i:i{-}$W$1$ colour space that we were unable to identify from spectroscopy. Further analysis with the J-band failed to prove conclusively that the unidentified targets in this colour space were stars. These targets exhibit broad band colours consistent with quasars. This could mean that the colour space is complete to fainter limits in $g$ than found in this work. We found that the combination of optical and mid-IR photometry improved the efficiency of our quasar selection. In Figure \ref{fig:spec_results} we see that fields where WISE photometry was included in the quasar selection saw an increased yield of ${\approx}10$ quasars deg$^{-2}$. The improvement in quasar selection that we found in this survey may readily be extended to other quasar surveys in a similar redshift range such as eBOSS. Furthermore, we note that WISE photometry has proven to be a boon for quasar selection at higher redshifts \citep{Carnall2015}.

\section{Clustering analysis}
\label{sec:qsoclustering}
\subsection{Correlation function estimators}
\subsubsection{Redshift space correlation function}
The two-point correlation function, $\xi(r)$, is commonly used to measure the excess probability of finding a pair of objects separated by distance $r$ over a pair of randomly distributed objects. This probability is given by \cite{Peebles1980} as;
\begin{equation}
dP=n^{2}[1+\xi(r)]dV_{1}dV_{2}
\end{equation}
where $n$ is the mean space density of objects and $dV$ are volume elements around object 1 and 2. When measuring quasar positions we measure their distribution in redshift space and so we recover $\xi(s)$ instead of $\xi(r)$. To recover an estimate for $\xi(s)$ we use the estimator of \cite{Landy1993};
\begin{equation}
\label{eqn:ls}
\xi _{LS}(s) = \frac{{<}DD{>} - 2{<}DR{>} + {<}RR{>}}{{<}RR{>}}
\end{equation}  
Here ${<}DD{>}$ is the number of quasar-quasar pairs at a given separation, denoted by $s$. The ${<}DR{>}$ and ${<}RR{>}$ terms correspond to the number of quasar-random and random-random pairs respectively. To reduce the Poisson noise, we calculate the DR and RR terms from a much larger (twenty times larger) sample of randoms than we have quasars. It is necessary to normalise these terms by the measured quasar density. As discussed in Section \ref{sec:stellar_density} our measured quasar density varies by as much as a factor of two over the sky as a result of variation in our selection function and observing conditions. We therefore normalise our random sample on a field by field basis dependent on the total number of quasars in a given field. This normalisation should help counter effects of photometric and spectroscopic incompleteness. Assuming quasar clustering to be described by a power-law with a correlation scale of $r_{0}{=}6.1 \: h^{-1}$Mpc and slope $\gamma{=}1.8$ the affect of the integral constraint from a single 2dF field is sufficiently small to have little effect on our measurements and only affects clustering measurements on the largest (${\approx}100 \: h^{-1}$Mpc) physical scales.
\subsubsection{Modelling quasar clustering in redshift space}
\label{sec:model}
Following the methodology of other quasar surveys (\cite{daAngela2005},\citetalias{daAngela2008}) we define $\xi(s)=\xi(\sigma^{2} + \pi^{2})$ where $\sigma$ is the pairwise separation perpendicular to the line of sight and $\pi$ is parallel to the line of sight and 
\begin{equation}
1 + \xi(\sigma,\pi) = \int_{-\infty}^{\infty}
[1+\xi'(\sigma,\pi-w_z(1+z)/H(z))]f(w_z)dw_z
\label{eqn:some_ref}
\end{equation}
\noindent and $\xi'(\sigma,\pi-w_z(1+z)/H(z))$ is given by equations 12-14 of \cite{daAngela2005}. These equations are completely equivalent to modelling the linear z-space distortions via the `Kaiser boost' of $\xi(s)=\xi(r (1+2/3\beta+1/5\beta^2))$ where $\beta(z)=\Omega_{m}(z)^{0.6}/b(z)$ is the infall parameter and $b(z)$ is the bias. $f(w_z)$ is the distribution of pairwise peculiar velocities, $w_z$, that includes the small-scale clustering motions of the quasars. From the above we can then derive $\xi(s)$ where $s=\sqrt{\sigma^2+\pi^2}$. Fitting $\xi(s)$ will form our main route to measuring quasar clustering amplitudes. We fit the correlation function between $5{<}s (h^{-1} \mathrm{Mpc}){<}55$ and assume a power-law model for $\xi(r)$ with $\xi(r)=(r/r_{0})^{-\gamma}$ and with a fixed $\gamma=1.8$.

At small scales ($s{\lesssim}5 h^{-1}$ Mpc) redshift space distortions dominate the clustering signal in $\xi(s)$. Non-linear,`finger-of-God' peculiar velocities of the quasars and redshift measurement errors both act to reduce $\xi(s)$ at these scales. Justified mainly by the good fit it provides, we shall assume a Gaussian for $f(w_z)$ \citep[see][]{Ratcliffe1996} with a fixed velocity dispersion of $\langle w_{z}^{2} \rangle ^{\frac{1}{2}} = 750 \: {\rm km s}^{-1}$.  

To fit $\xi(s)$ we need an initial model for quasar bias and its dependence on redshift. We shall assume the previous 2QZ fit of \citetalias{Croom2005}; 
\begin{equation}
b = 0.53+0.289(1+z)^2.
\label{eqn:empbias}
\end{equation}
This implies a $\pm10$\% difference to $1+2/3\beta(z)$ in the $0.5<z<2.5$ range wrt. the median redshift $z{=}1.4$. This corresponds to a $\pm5\%$ effect in $r_{0}$. Therefore, we cannot assume that $\beta$ is independent of redshift in fitting $\xi(s)$ for $r_{0}$. We fit $\xi(s)$ using the above model for bias and we shall check for consistency with our new results for the $z$ dependence of bias at the end of our analysis.

\subsection{2QDESp Quasar correlation function}
\label{sec:2pcf}
\begin{figure}
\centering
\includegraphics[width=\linewidth]{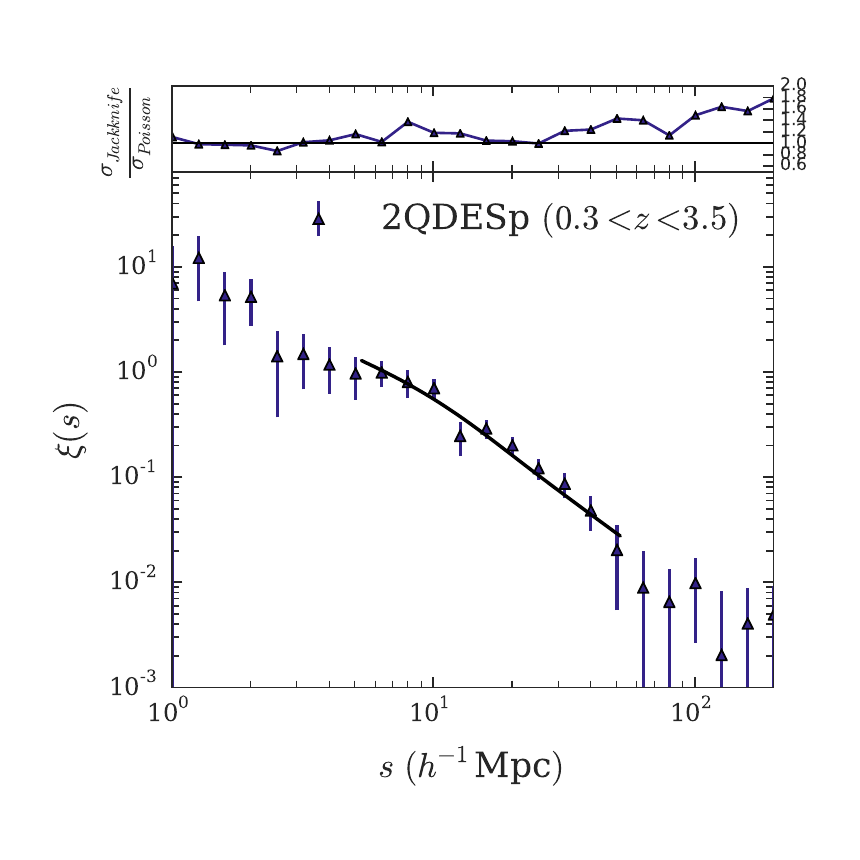}
\caption{We show the measured $\xi(s)$ for the 2QDESp quasar sample between $0.3{<}z{<}3.5$. We include our model with best-fitting correlation length (using Jackknife errors) of $r_{0}{=}6.25{\pm}0.30 h^{-1}$Mpc. In the top panel, we show the ratio between the Jackknife and Poisson errors.}
\label{fig:2qdes_xis}
\end{figure}

We present the $z$-space two point correlation function, $\xi(s)$, measured from the 2QDESp sample for $0.3{\leq}z{\leq}3.5$ in Figure \ref{fig:2qdes_xis}. Widening the redshift interval from our targeted redshift range ($0.8{<}z{<}2.5$) maximises the signal of the correlation function. We have considered two estimates of the errors, Poisson and jackknife. At small enough scales and sparse enough space densities, it is well known that the errors in $\xi$ can be approximated by Poisson errors. Usually measured as $\Delta\xi(s)=1+\xi(s)/\sqrt{<DD>}$, which is the error assuming that the clustering signal is zero. However, in cases where value of $\xi(s)$ is negative, this estimate under predicts the error. In these cases we model the Poisson error to be;
\begin{equation}
\Delta \xi(s) = \sqrt{\frac{1+\xi(s)_{\mathrm{MODEL}}}{<DR>_{\mathrm{observed}}}}
\label{eqn:modelerr}
\end{equation}
When a bin in $s$ is well populated with quasar-quasar pairs this error estimate and the Poisson error approximately converge. The error estimate of equation \ref{eqn:modelerr} is used only within Section \ref{sec:absmagred} due to the $s$ bins being more sparsely sampled. We note that this error appears insensitive to a range of model $r_{0}$ values and is instead sensitive to the $<DR>_{\mathrm{observed}}$ value.

At these smaller scales, the covariance between $\xi$ estimates may be ignored in fits on the basis that the pair counts are close to being independent. We demonstrate his by comparing jackknife and Poisson errors for 2QDESp. To calculate the jackknife errors, we split the data into ${\approx} 60$ subsamples (each separate 2dF pointing) and calculate the error using a jackknife approach described by:
\begin{equation}
\sigma_{jackknife} =
\sqrt{\sum_{i=1}^{N}\frac{DR_{i}(s)}{DR_{total}(s)}(\xi_{i}(s)-\xi_{total}(s))^{2}},
\label{eqn:jackknife}
\end{equation}
where $N$ is the total number of subsamples, the $i$ subscript denotes which 2dF field has been removed from the whole sample and the total subscript denotes use of the full sample. Here we weight each term within the sum by the number of data-random pairs excluded from the calculation and so weight more densely sampled fields more highly than those with fewer data random pairs. 

For comparison, we show the ratio of the jackknife error estimation to Poisson errors in the top panel of Figure \ref{fig:2qdes_xis}. We see that the Poisson error is a reasonable representation of the jackknife error out to $s{<}30 \: h^{-1}$Mpc and is still within a factor of ${\approx}1.4$ at $s{<}55 \: h^{-1}$Mpc, only reaching a factor of ${\approx}1.8$ at $s{\approx} 100{-}200 \: h^{-1}$Mpc. This suggests that, at least out to $s{<}55 \: h^{-1}$Mpc, pair counts are reasonably independent and this is supported by the small size of off-diagonal terms in the covariance matrix at these scales. We shall fit models in the range $5{<}s{<}55 \: h^{-1}$Mpc using both jackknife and Poisson errors.

We now fit the model from Section \ref{sec:model} to the 2QDESp $\xi(s)$ data. We show our best-fitting model assuming Poisson errors in the lower panel of Figure \ref{fig:2qdes_xis}. This has $r_{0}=6.25{\pm}0.25 \: h^{-1}$Mpc which fits well with $\chi^2=9.4$ with 10 degrees-of-freedom, (df.). Assuming jackknife errors for the fit gives a similar result, $r_{0}=6.25{\pm}0.30\: h^{-1}$Mpc ($\chi^{2},$df.$=7.0,10$).

\begin{figure}
\centering
\includegraphics[width=\linewidth]{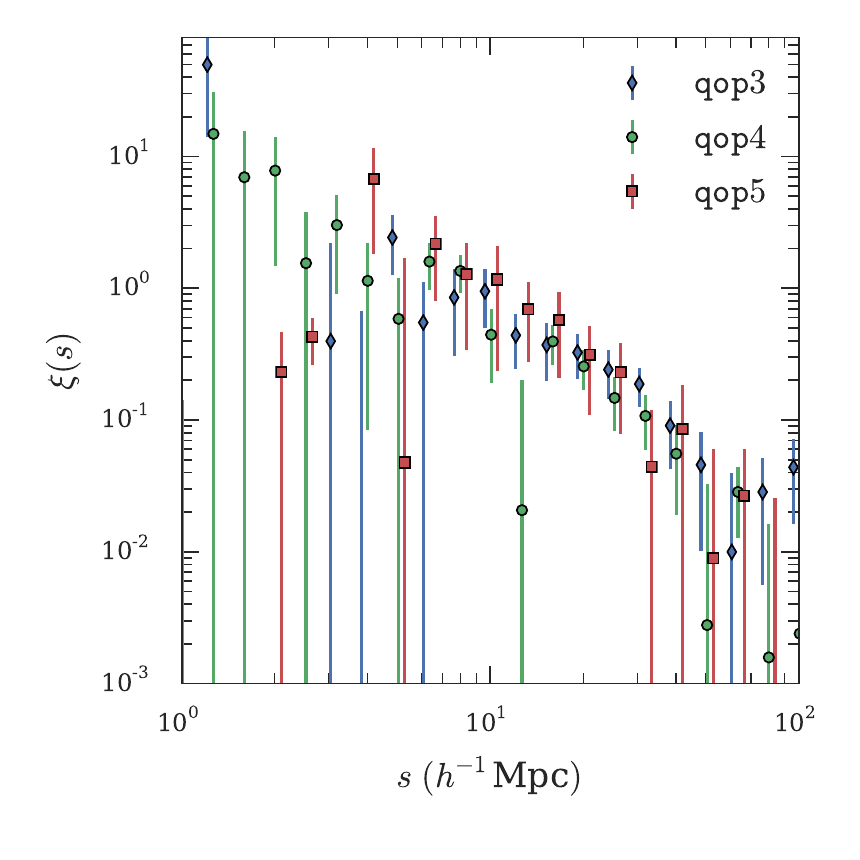}
\caption{Here we show the correlation function for 2QDESp quasars
with the highest, intermediate and lowest quality spectra, {\tt qop} = 5,4 and 3 respectively. We offset the high and low spectral quality correlation functions along the x-axis by $10^{s\pm0.02}$ for clarity. The three correlation functions for each quality level agree. Hence we argue that the lowest quality sample is suitable for use in our analysis.} 
\label{fig:qop}
\end{figure}

To further test the quality of our data, we divide the quasar sample into three subsets, based on the quality of their optical spectra. The three subsets consist of $1675$, $4585$ and $3541$ quasars for the highest, intermediate and lowest quality spectra respectively. We compare the different quality spectra in Figure \ref{fig:qop}, where we plot the correlation function for the three quasar subsamples. We fit the model from Section \ref{sec:model} to each and find that the three subsamples agree at the ${\sim} 1 \sigma$ level. Using this procedure we verify that our lowest quality optical spectra are suitable to use in our science measurements as the contamination by other sources is low enough not to cause significant differences in the measured clustering signal.

\begin{table*}
\hbox{\hspace{-100ex}}
\resizebox{2\columnwidth}{!}{%
\begin{tabular}{lllllll}
\hline
\hline
Survey & $r_{0} (h^{-1}$Mpc$)$ & Faint limit & Median mag. & Median $z$ & N$_{QSO}$&$\chi^{2} (10 df.) $\\
       & $(\gamma=1.8)$       &             &      $(g)$       &            &          & ($r_{0}=6.1 h^{-1}$Mpc) \\
\hline
SDSS   & $6.55^{+0.30}_{-0.30}$     & $g{<}19.4$              & $18.8$ & $1.37$ &  $32 650$& $4.7$\\
2QZ    & $5.85^{+0.20}_{-0.20}$     & $g{<}20.8$              & $20.1$ & $1.48$ &  $22 211$& $14.9$\\
2QDESp  & $6.10^{+0.25}_{-0.25}$     & $g{<}22.5$              & $20.6$ & $1.54$ &  $9 705$ & $12.1$\\
2SLAQ  & $6.15^{+0.35}_{-0.35}$     & $g{<}21.9$              & $21.3$ & $1.58$ &  $6 374$ & $15.6$\\
\end{tabular}%
}
\caption{We present model fits for the re-analysed data sets, 2QZ, 2SLAQ and SDSS DR5 as well as for the 2QDESp sample. We restrict our analysis to quasars between $0.3 < z < 2.9$ to ensure good agreement between the redshift distributions. We include the best-fitting $r_{0}$, the faint limits of the quasar samples as well as their median magnitudes, redshifts and number of quasars. We note that limiting our analysis (in the case of 2QDESp) to this redshift interval changes the best fitting value compared to Section \protect \ref{sec:2pcf}. However, this change is $<1\sigma$ and is discussed in Section \protect \ref{sec:reddep}.}
\label{tab:other_surveys}
\end{table*}
\begin{figure*}
\centering
\hspace{-0.3cm}
\includegraphics[width=\linewidth]{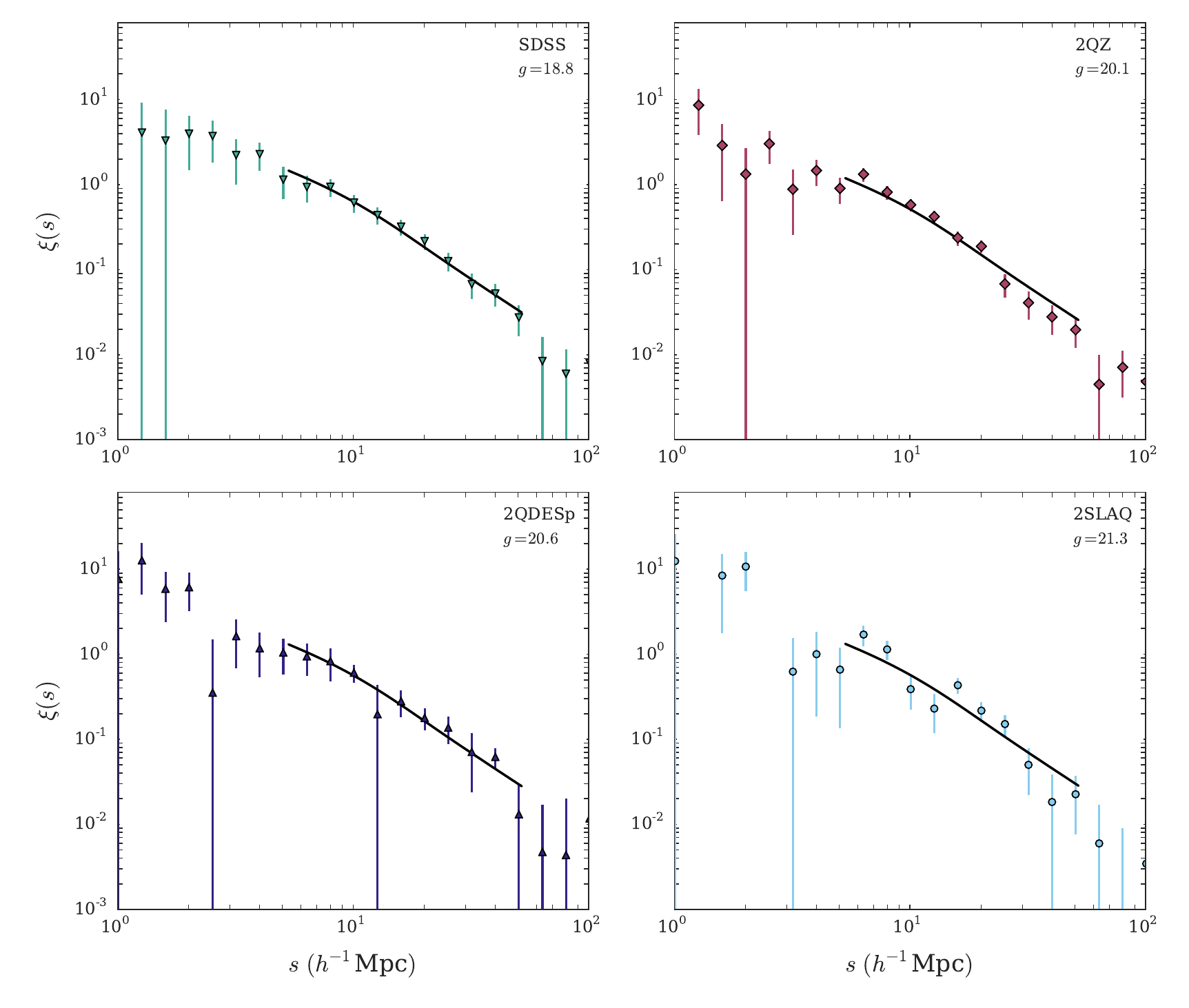}
\caption{Each panel shows our estimate of $\xi(s)$ measured for a particular wide area survey as labelled. We annotate each panel with the median magnitude in $g$ for comparison to our survey. Errors are Poisson. We fit the data using the model from section \ref{sec:model}, where we assume Gaussian velocity dispersions in real space, $\gamma = 1.8$ with a velocity dispersion, $\langle\omega^{2}\rangle^{\frac{1}{2}} = 750 \: \mathrm{km} \: s^{-1}$. In each panel we show the model where $r_{0}{=}6.1\:h^{-1}$Mpc (solid line) (see section \protect \ref{sec:model}). For each survey we restrict the analysis to the redshift interval $0.3{<}z{<}2.9$ as this range is well sampled by all surveys. The best-fitting models for the individual surveys are shown in Table \ref{tab:other_surveys}.}
\label{fig:four_surveys}
\end{figure*}
\subsection{Luminosity dependence of clustering}
\label{sec:ldep}
In this section we search for evidence of luminosity dependent quasar clustering. We start with the approach of \citetalias{Shanks2011} and compare measured $r_{0}$ values between different surveys at approximately fixed redshift. We follow this with the more precise methodology of \citetalias{daAngela2008} which divides the samples by absolute magnitude and redshift. We defer measurement of redshift dependence to Section \ref{sec:reddep}.

\subsubsection{Apparent magnitude}
\label{sec:apparent}
Comparison between the 2QZ, SDSS, 2SLAQ and 2QDESp quasar surveys provides an opportunity to measure the dependence of clustering on luminosity. Whilst each survey has different selection methods and flux limits we see in Figure \ref{fig:nz} that the resulting redshift distributions are similar (see also Table \ref{tab:other_surveys}). Given that each survey is flux limited, we account for photometric and spectroscopic incompletenesses by characterising each survey by its median magnitude. For the four surveys this corresponds to; $g{=}18.8$ (SDSS), $g{=}20.1$ (2QZ, see \cite{Richards2005} for $b_{J}-g$ conversion), $g{=}21.3$ (2SLAQ) and $g{=}20.6$ for the 2QDESp sample.

In Figure \ref{fig:four_surveys} we show our re-analysis of the SDSS, 2QZ, 2SLAQ and 2QDESp clustering results, restricting our analysis to quasars between $0.3{<}z{<}2.9$. We have rebinned the $s$-axis to a common binning across the four surveys. In each panel we show the best-fit $r_{0}$ for each survey, assuming a fixed $\beta$. We permit use of constant $\beta$ here because with the small difference in median redshifts the effect of different $\beta(z)$ values will have $<1\%$ effect on $r_{0}$. Our best-fitting values are shown in Table \ref{tab:other_surveys} and these measurements agree with the analysis of \citetalias{Shanks2011} with any differences in the best-fit values due to slight difference in redshift range and fitting interval. 

\begin{figure}
\centering
\includegraphics[width=\linewidth]{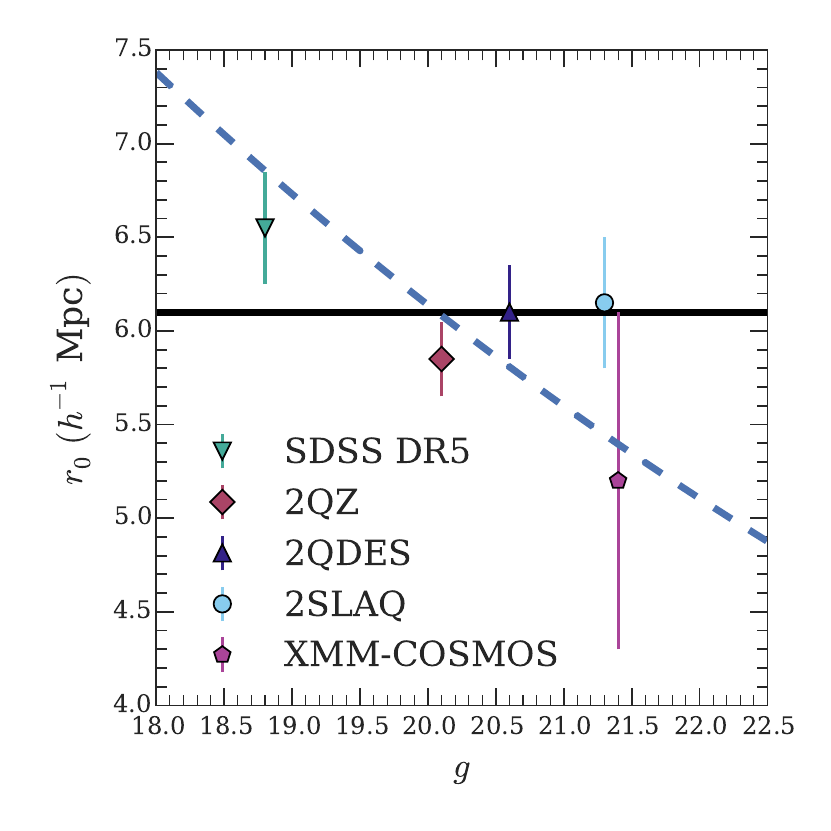}
\caption{We show the median depth for 2QZ, 2SLAQ, SDSS and 2QDESp surveys along with the best-fit $r_{0}$ with the associated errors. We also show a flat $r_{0}=6.1 \: h^{-1}$ Mpc model (solid line) and a $L^{0.1}$ model (dotted line).}
\label{fig:foo}
\end{figure}

We make a comparison between the median magnitude and best-fit values of $r_{0}$ across the four surveys in Figure \ref{fig:foo}. We note that \cite{Shen2009} found that the brightest SDSS DR5 quasars clustered more strongly than the rest of their quasar sample. We find here that $r_{0}$ for the SDSS quasars is larger than the $r_{0}$ values from the other surveys but only at ${\approx}1\sigma$ level. As this effect corresponds to the result reported by \cite{Shen2009} we must be cautious not to immediately dismiss the difference as purely statistical. However, we further test for the dependence on $r_{0}$ with magnitude using the Spearman rank correlation test. We find a Spearman rank order correlation of $\rho{=}-0.19\pm0.37$ which is consistent with no correlation between apparent magnitude and clustering scale. We also find that the points in Figure \ref{fig:foo} are consistent with a constant $r_{0} = 6.10^{+0.10}_{-0.10}\:h^{-1}$Mpc with $\chi^2,$df.$=3.9,3$ and p-value$=0.28$.

In Table \ref{tab:other_surveys} we calculate the corresponding $\chi^{2}$ when we compare each survey individually to a fixed $r_{0} = 6.1 h^{-1}$Mpc and we find that the total $\chi^{2},$df.$=46.8,40$ (we include the individual survey $\chi^{2}$ values in Table \ref{tab:other_surveys}). So from this analysis we are unable to reject the hypothesis that quasar clustering is independent of luminosity from a comparison between the individual surveys.
\subsubsection{Absolute magnitude}
\label{sec:absmagred}
In Section \ref{sec:apparent} we compared quasar clustering over a range of ${\sim}3.5$ magnitudes at fixed redshift. Although further subdivision of the quasar samples will yield weaker statistical constraints, we are, however, able to probe a much larger dynamic range ($-22.3<M_{i}(z=2)<-28.5$) at fixed redshift by combining all four surveys. We do this by taking the error weighted mean of the four correlation functions for each subsample. Following the approach of \citetalias{daAngela2008}, we divide the M-$z$ plane into non-overlapping bins. We use the sample binning of \citetalias{daAngela2008} which was designed to maximise the clustering signal from the 2QZ$+$2SLAQ combined sample. The inclusion of the SDSS and 2QDESp data reduces the statistical errors; particularly in the highest and lowest luminosity bins. This may enable us to potentially uncover previously hidden dependencies. 

We therefore subdivide the quasar samples into thirteen, non-overlapping subsets in luminosity and redshift. The absence of low-luminosity quasars at high redshift limits the dynamic range (in luminosity) at the higher redshifts. We calculate the absolute magnitudes $M_{i}$ using
\begin{equation}
M_{i} = i - A_{i} - 25 - 5log(d) - K_{i},
\label{eqn:absmag}
\end{equation}
where $i$ is the apparent magnitude, $A_{i}$ is the dust extinction, $d$ is the luminosity distance in Mpc and $K_{i}$ is the k-correction in the $i$-band. The galactic dust correction, $A_{i}$ is calculated by $A_{i}=2.086\ E(B-V)$ \citep{Schlegel1998}. The k-correction value was taken from \cite{Richards2006}. 

\begin{figure*}
\centering
\includegraphics[width=\textwidth]{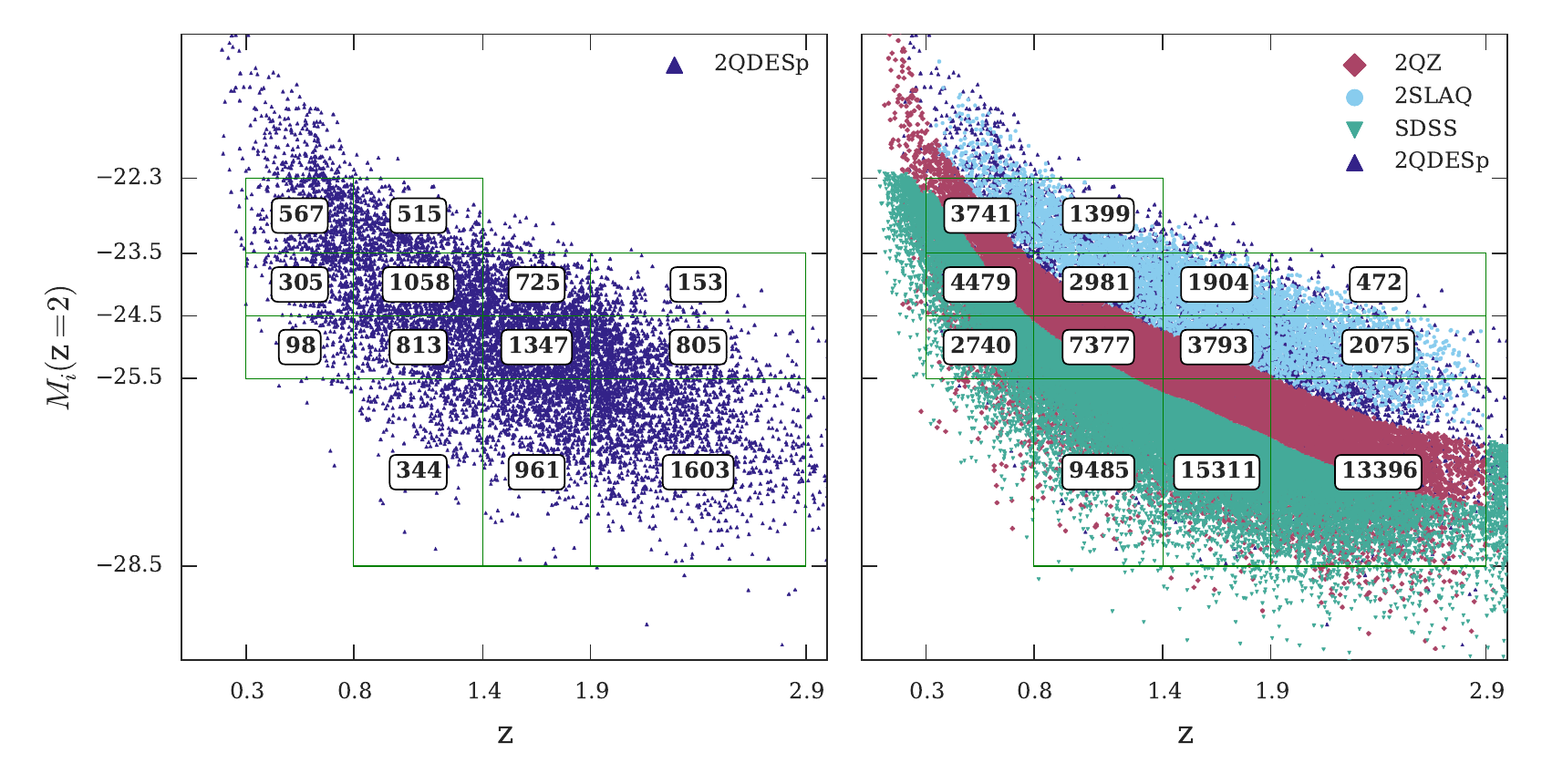}
\caption{The distribution of our sample in redshift-luminosity (left) and the comparison to 2QZ, 2SLAQ, SDSS DR5 and 2QDESp surveys (right). The grids show the division in magnitude and redshift applied to the samples and the occupancy of each bin.}
\label{fig:absmag_vs_redshift}
\end{figure*}

We show the distribution of the 2QDESp sample in the left hand panel of Figure \ref{fig:absmag_vs_redshift}. We overlay the Figure with the M-$z$ divisions and include the occupancy of each division. In the right panel of Figure \ref{fig:absmag_vs_redshift} we plot the M-$z$ distribution for the combined (SDSS$+$2QZ$+$2SLAQ$+$2QDESp) sample. Again, we overlay the M-$z$ divisions and show the total bin occupancy. In both panels of Figure \ref{fig:absmag_vs_redshift} the flux limited nature of the surveys is evident by the absence of lower luminosity quasars at higher redshifts. As expected, we see that the 2QDESp survey makes its largest contribution at fainter absolute luminosities. 

\begin{figure*}
\centering
\includegraphics[width=\textwidth]{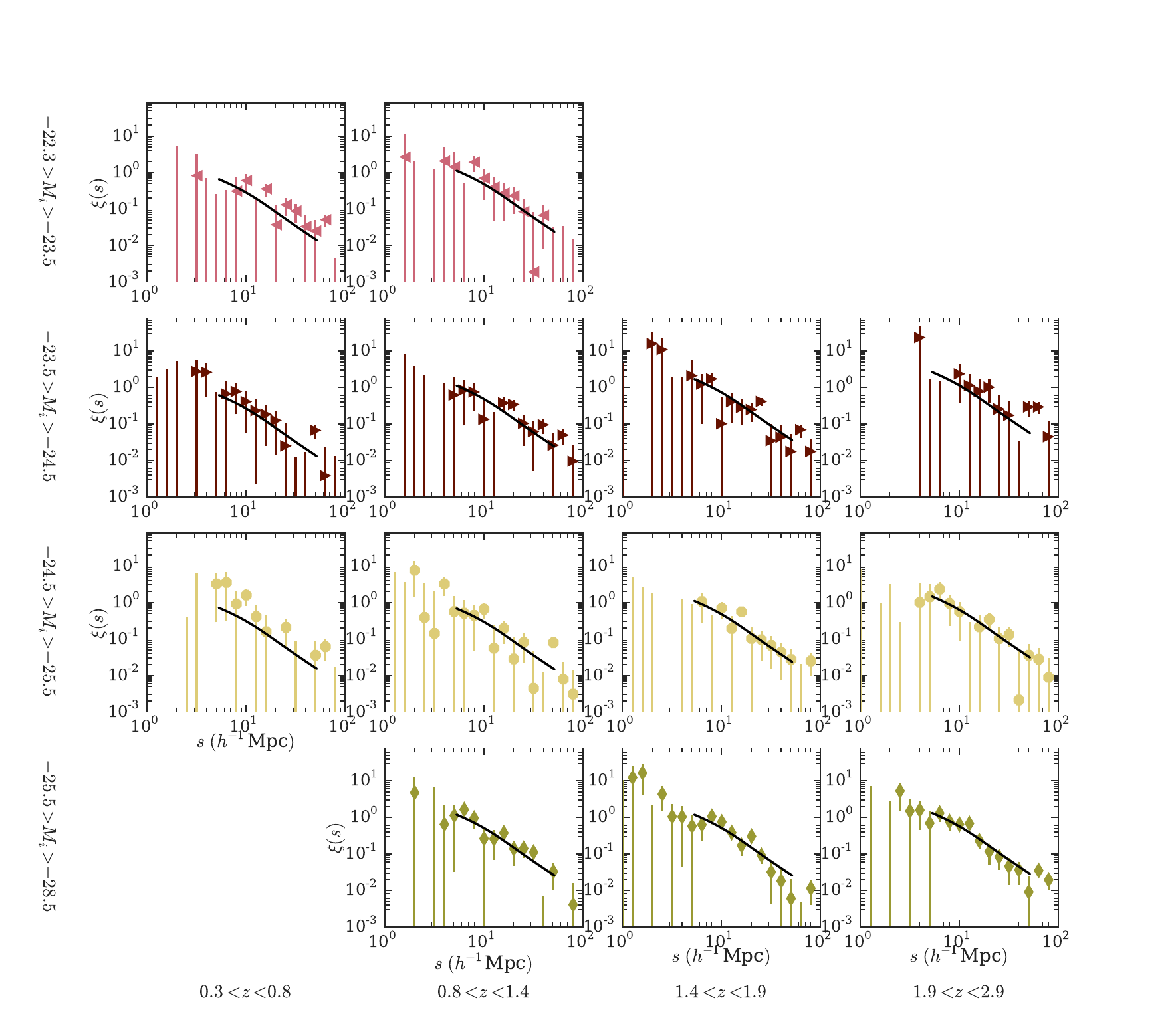}
\caption{We measure the correlation function $\xi(s)$ for the combined sample (SDSS,2QZ, 2SLAQ and 2QDESp) in the same bins as Figure \protect{\ref{fig:absmag_vs_redshift}}. We use the error weighted mean to combine the measurements from each individual survey, where errors are Poisson (see Section \protect \ref{sec:2pcf}). These are compared to a $\xi(s)$ model where $r_{0}=6.1 h^{-1}$ Mpc (solid line). We show the fit quality for this fixed $r_{0}$ value as well as for the best-fitting value in Table \protect{\ref{tab:absolute_redshift_table}}.}
\label{fig:mag_red_cuts_whole}
\end{figure*}

Our aim in subdividing the combined sample in both magnitude and in redshift is to isolate redshift and luminosity dependent effects on the clustering amplitude. In Figure \ref{fig:mag_red_cuts_whole} we show the signal for each of the absolute magnitude and redshift bins.
\begin{table*}
\resizebox{2\columnwidth}{!}{%
\begin{tabular}{cccccccccc}
\hline
\hline
Redshift range & $z$ & Absolute magnitude range & $M_{i}(z=2)$ & Best $r_{0}$ & $\chi^{2}$ & p-value & $\xi_{20}$ & b & $M_{DM} \times 10^{12} h^{-1} \mathrm{M_{\odot}}$\\
 & (median) & $M_{i}(z=2)$ & (median) & & &  \\
\hline

$0.3<z<0.8$ & $0.53$ & $-23.5<Mi<-22.3$ & $-23.06$ & $4.2^{+0.65}_{-0.8}$ & $11.96$ & $0.29$ &$ 0.18\pm 0.06$ & $-$             & $-$                      \\
$0.8<z<1.4$ & $0.99$ & $-23.5<Mi<-22.3$ & $-23.18$ & $5.65^{+1.0}_{-1.2}$ & $5.95$ & $0.82$  &$ 0.39\pm 0.11$ & $ 1.79\pm 0.29$ & $ 3.60_{- 2.19}^{+ 3.54}$\\
$0.3<z<0.8$ & $0.63$ & $-24.5<Mi<-23.5$ & $-23.93$ & $4.05^{+0.8}_{-1.05}$ & $8.7$ & $0.56$  &$ 0.25\pm 0.08$ & $ 1.13\pm 0.21$ & $ 0.71_{- 0.62}^{+ 1.60}$\\
$0.8<z<1.4$ & $1.10$ & $-24.5<Mi<-23.5$ & $-24.04$ & $5.65^{+0.7}_{-0.8}$ & $8.91$ & $0.54$  &$ 0.29\pm 0.08$ & $ 1.58\pm 0.24$ & $ 1.40_{- 0.90}^{+ 1.55}$\\
$1.4<z<1.9$ & $1.61$ & $-24.5<Mi<-23.5$ & $-24.15$ & $7.1^{+0.8}_{-0.9}$ & $10.77$ & $0.38$  &$ 0.40\pm 0.10$ & $ 2.39\pm 0.33$ & $ 2.87_{- 1.43}^{+ 2.08}$\\
$1.9<z<2.9$ & $2.02$ & $-24.5<Mi<-23.5$ & $-24.31$ & $9.1^{+2.7}_{-3.65}$ & $9.26$ & $0.51$  &$ 1.08\pm 0.43$ & $ 4.79\pm 0.96$ & $17.17_{- 9.23}^{+13.35}$\\
$0.3<z<0.8$ & $0.69$ & $-25.5<Mi<-24.5$ & $-24.84$ & $4.4^{+1.5}_{-2.1}$ & $13.0$ & $0.22$   &$ 0.37\pm 0.15$ & $ 1.51\pm 0.34$ & $ 3.62_{- 2.92}^{+ 6.18}$\\
$0.8<z<1.4$ & $1.07$ & $-25.5<Mi<-24.5$ & $-25.09$ & $4.3^{+0.7}_{-0.75}$ & $13.98$ & $0.17$ &$ 0.21\pm 0.06$ & $ 1.29\pm 0.21$ & $ 0.42_{- 0.33}^{+ 0.72}$\\
$1.4<z<1.9$ & $1.65$ & $-25.5<Mi<-24.5$ & $-24.99$ & $5.6^{+0.7}_{-0.85}$ & $8.84$ & $0.55$  &$ 0.36\pm 0.08$ & $ 2.28\pm 0.27$ & $ 2.10_{- 0.96}^{+ 1.35}$\\
$1.9<z<2.9$ & $2.11$ & $-25.5<Mi<-24.5$ & $-25.07$ & $6.55^{+0.95}_{-1.1}$ & $5.52$ & $0.85$ &$ 0.33\pm 0.11$ & $ 2.60\pm 0.44$ & $ 1.36_{- 0.81}^{+ 1.33}$\\
$0.8<z<1.4$ & $1.21$ & $-28.5<Mi<-25.5$ & $-25.94$ & $5.85^{+0.6}_{-0.6}$ & $9.08$ & $0.52$  &$ 0.34\pm 0.07$ & $ 1.84\pm 0.21$ & $ 2.27_{- 1.06}^{+ 1.52}$\\
$1.4<z<1.9$ & $1.67$ & $-28.5<Mi<-25.5$ & $-26.27$ & $5.85^{+0.35}_{-0.4}$ & $16.98$ & $0.07$&$ 0.37\pm 0.04$ & $ 2.35\pm 0.15$ & $ 2.28_{- 0.59}^{+ 0.71}$\\
$1.9<z<2.9$ & $2.19$ & $-28.5<Mi<-25.5$ & $-26.58$ & $6.2^{+0.45}_{-0.5}$ & $7.63$ & $0.67$  &$ 0.39\pm 0.05$ & $ 2.91\pm 0.20$ & $ 1.90_{- 0.51}^{+ 0.62}$\\

\end{tabular}%
}
\caption{We show the best-fit value of $r_{0}$ for each $M-z$ bin with the corresponding error, $\chi^{2}$ and p-value. We correct for varying $\beta(z)$ according to equation \protect{\ref{eqn:empbias}}. We fit between $5{<}s \:(h^{-1}$ Mpc$){<}55$, each bin having $10$ df.. We include measurements of $\xi_{20}$ (section \protect{\ref{sec:reddep}}), bias and dark matter halo mass (section \protect{\ref{sec:bias}}).}
\label{tab:absolute_redshift_table}
\end{table*}
To generate random samples we use R.A.-Dec. mixing \citepalias[see][]{Croom2005}, sampling from the all magnitudes and redshifts to generate the angular mask. The radial mask is generated by randomly sampling the redshift distribution of the magnitude-redshift subsample. We found that fitting the radial distribution with a polynomial provided similar results to those included here.

Previously we fit for $r_{0}$ at approximately fixed redshift. However, here we are fitting over $\Delta z {\sim} 1.7$ and so an assumption of constant $\beta$ is no longer valid. We therefore measure the correlation function for each subsample in Figure \ref{fig:mag_red_cuts_whole} but determining $\beta(z)$ from an assumed b($z$) relationship (see equation \ref{eqn:empbias}). Whilst there is uncertainty in the precise form of the b($z$), a $50\%$ increase in bias at $z=1.5$ only results in a $4\%$ change in $r_{0}$.
\begin{figure}
\hbox{\hspace{1cm}}
\includegraphics[width=\linewidth]{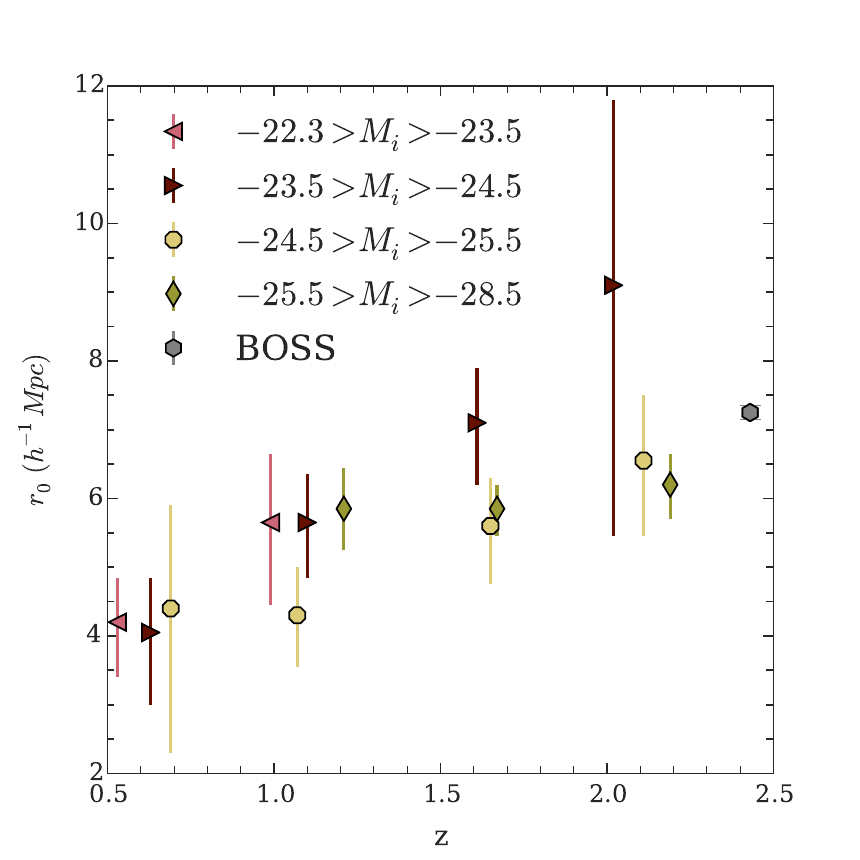}
\caption{We show the measured correlation length ($r_{0}$) for the thirteen luminosity$-$redshift bins from Table \protect{\ref{tab:absolute_redshift_table}}. We include our measurement of $r_{0}$ from the correlation function of 
\protect \cite{Eftekharzadeh2015} who measure the clustering scale of quasars from the BOSS sample as $r_{0}=7.25 \pm 0.10 \: h^{-1}$ Mpc.}
\label{fig:r0_vs_z}
\end{figure}

Allowing the value of $r_{0}$ to vary between each bin we find a total $\chi^{2},$df.$=130.6,130$ and p-value$=0.47$. We plot the best-fit values (see Table \ref{tab:absolute_redshift_table}) in Figure \ref{fig:r0_vs_z}. We also include in this Figure our determination of $r_{0}$ from the measured $\xi(s)$ of \cite{Eftekharzadeh2015}; using our model we find their correlation function corresponds to $r_{0}{=}7.25{\pm}0.1 \: h^{-1}$ Mpc. 

In Figure \ref{fig:r0_vs_z} we compare across the luminosity bins at approximately fixed redshift. The fainter two magnitude bins (spanning $-24.5{<}M_{i}{<}-22.3$) show, on average, stronger clustering at all redshifts than the brighter bins. If this trend is physical it would suggest that fainter quasars are more strongly clustered than brighter quasars, suggesting an inverse relationship between quasar luminosity and halo mass.  However, we note that these magnitude-redshift bins correspond to the faintest apparent magnitudes in the 2QDESp and 2SLAQ samples and suffer from large incompleteness. So although there may be a weak underlying dependence on luminosity we are unable to claim a significant detection analysing the data in this fashion. It is possible, of course, that some effect of luminosity dependence is being masked by the redshift dependence of quasar clustering.

\subsection{Redshift dependence}
\label{sec:reddep}
In Figure \ref{fig:r0_vs_z} we see evidence for redshift dependence of quasar clustering and find that the increase in $r_{0}$ with redshift is significantly detected using the Spearman rank order correlation test ($\rho=0.82\pm0.18$). Here we attempt to measure the evolution of quasar clustering with redshift. Following the methodology of earlier authors \citepalias{Croom2005,daAngela2008} we use the integrated correlation function. We measure the clustering excess up to some radius ($s<20\: h^{-1} Mpc$) and normalise the signal according to the average quasar numbers contained within a $20\: h^{-1} Mpc$ radius sphere;
\begin{equation}
\label{eqn:xi20}
\xi_{20} = \frac{3}{20^{3}}\int^{20}_{0} \xi(s)s^{2} ds
\end{equation}
\begin{figure}
\centering
\includegraphics[width=\linewidth]{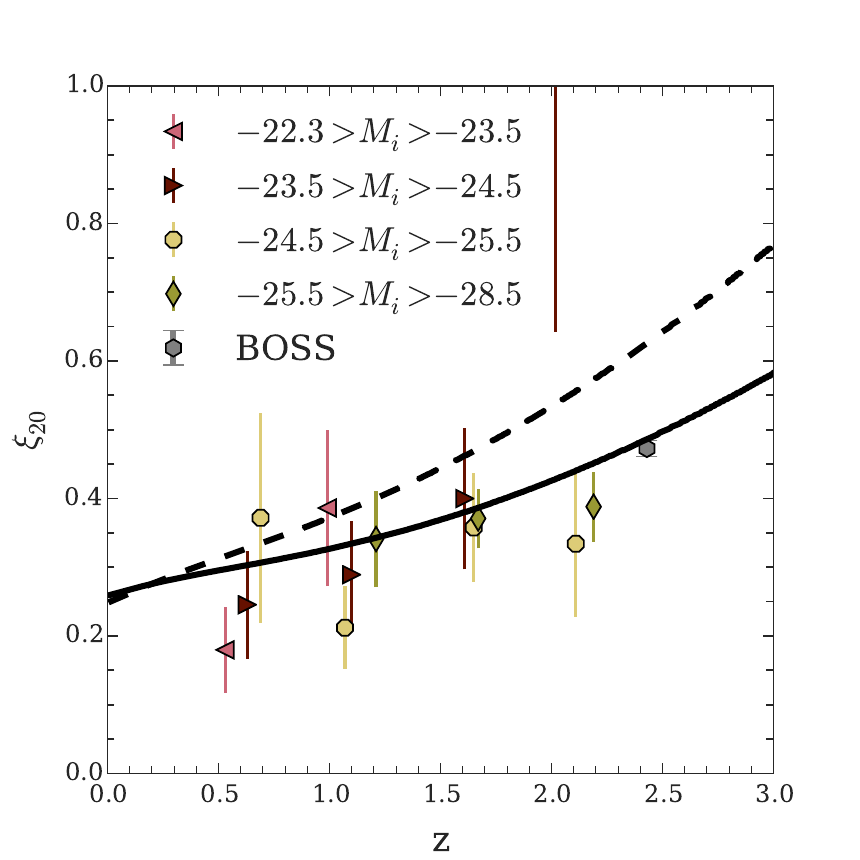}
\caption{ We show the measured $\xi_{20}^{Q}$ for the bins from Section \protect{\ref{sec:absmagred}}. We include model predictions for the evolution with redshift of $\xi_{20}^{Q}$. The solid line shows the expected $\xi_{20}^{Q}(z)$ relation assuming the empirical $b(z)$ relationship from equation \ref{eqn:newempbias}. For comparison we show the empirical $b(z)$ relation from \protect \citetalias{Croom2005} as a dashed line, i.e. equation \ref{eqn:empbias}.}
\label{fig:xitwenty}
\end{figure}
\citetalias{Croom2005} in particular looked at the effect of both systematic and statistical uncertainties associated with integrating different radius spheres. We adopt the same radius as used by these authors \citepalias[see][for a detailed analysis]{Croom2005}. 

In Figure \ref{fig:xitwenty} we show the integrated correlation function for each absolute magnitude and redshift bin from Section \ref{sec:absmagred}. We show the redshifts and $\xi_{20}$ values for these bins in Table \ref{tab:absolute_redshift_table}. We see that the evolution of $\xi_{20}(z)$ is flatter than one might naively expect from either Table \ref{tab:absolute_redshift_table} or Figure \ref{fig:r0_vs_z}. This is due to the effect of $\beta(z)$, accounted for in our model, that ``boosts'' $\xi_{20}(z)$ more at lower redshifts than at higher redshifts and thus flattens the evolution of $\xi_{20}$.

\subsubsection{Bias \& Halo Masses}
\label{sec:bias}
2QZ measured the quasar correlation function as a function of redshift \citepalias[see][]{Croom2005}. They reported the relationship of quasar bias with redshift described by equation \ref{eqn:empbias}. In this section we use the same methodology as previous works (\citetalias{Croom2005};\citetalias{daAngela2008} and \citetalias{Ross2009}) with our larger dataset to more precisely determine the evolution of bias with redshift.

We assume a scale independent bias and thus obtain;

\begin{equation}
b=\sqrt{\frac{\xi^{Q}(r)}{\xi^{\rho}(r)}} {\simeq} \sqrt{\frac{\xi_{20}^{Q}}{\xi_{20}^{\rho}}},
\label{eqn:bias}
\end{equation}
where $\xi^{Q}(r)$ and $\xi^{\rho}(r)$ are the quasar and matter real space correlation functions, with $\xi_{20}^{Q}$ and $\xi_{20}^{\rho}$ being the corresponding integrated correlation functions to $s<20 h^{-1}$ Mpc. \cite{Kaiser1987} describes the relation between the real and $z-$space correlation functions on linear scales as

\begin{equation}
\xi_{20}^{Q}(s) = \Big( 1+\frac{2}{3}\beta+\frac{1}{5}\beta^{2} \Big) \xi_{20}^{Q}(r).
\end{equation}

\noindent This results in an expression for quasar bias as a function of redshift;

\begin{equation}
b(z) = \sqrt{\frac{\xi_{20}^{Q}(s)}{\xi_{20}^{\rho}(r)}-\frac{4\Omega_{m}^{1.2}(z)}{45}}-\frac{\Omega_{m}^{0.6}}{3}.
\end{equation}
In line with earlier work we use $0.6$ as the exponent to $\Omega_{m}$. To estimate $\xi_{20}^{\rho}(r)$ we use the matter power-spectrum at $z=0$. This was calculated using {\tt CAMB } \citep{Lewis2000,Challinor2011}, which is based on {\tt CMBFAST } \citep{Seljak1996,Zaldarriaga2000}. Under our assumed cosmology we find $\xi_{20}^{\rho}(r)=0.253$ at $z=0$. We can then use linear theory to convert from a measured $\xi_{20}^{Q}$ to bias ($b$) via equation \ref{eqn:bias}. We correct for non-linear effects in the same manner as described by \citetalias{Croom2005}.

\begin{figure}
\centering
\includegraphics[width=\linewidth]{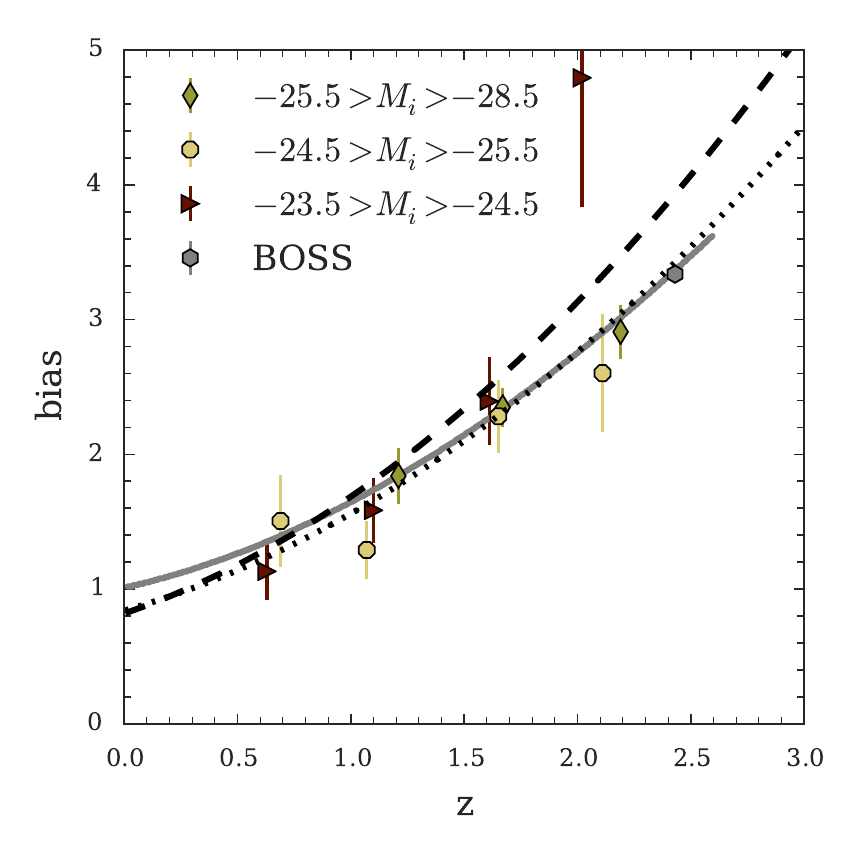}
\caption{We show our estimate of quasar bias as a function of $z$ and absolute magnitude. We include our measurement of bias from the BOSS survey, \protect \cite{Eftekharzadeh2015}. We show the evolution for a halo of mass $2\times 10^{12} \: h^{-1} \: \mathrm{M_{\odot}}$ as the solid grey line. We see that our measurements of bias are consistent with quasars inhabiting the same mass halos irrespective of magnitude or redshift. We include the 2QZ bias result (Equation \protect \ref{eqn:empbias}) as a black dashed line and our bias result (Equation \protect \ref{eqn:newempbias}) as a dotted black line for comparison.}
\label{fig:bias}
\end{figure}

Figure \ref{fig:bias} shows how the resulting bias varies with $z$. We fit an empirical relationship to the results in Figure\ref{fig:bias};
\begin{equation}
b(z) = (0.59\pm0.19) \:+\: (0.23\pm0.02)(1+z)^{2}.
\label{eqn:newempbias}
\end{equation}
We note that this $z$ dependence has the same quadratic form than that of equation \ref{eqn:empbias} but with a weaker gradient. We refer back to Section \ref{sec:absmagred} where we discussed the effect of different $b(z)$ models on the measurement of $r_{0}$. We remeasure the $r_{0}$ fits from earlier sections and find changes in the best-fit values are of the order $\pm0.05 \: h^{-1}$ Mpc, well below our statistical error.

Figure \ref{fig:xitwenty} shows the difference the change in the b$(z)$ relationship makes on $\xi_{20}$. The dashed line showing the prediction of $\xi_{20}$ from $\xi_{\rho}(r,z=0)$ and equation \ref{eqn:empbias} and the solid line showing the prediction of equation \ref{eqn:newempbias}. We also plot the independent BOSS data from \cite{Eftekharzadeh2015} which lies much closer to our b($z$) result than that of 2QZ.

Having derived bias values from our measured values of $\xi_{20}^{Q}(s)$ (see Table \ref{tab:absolute_redshift_table}) we want to relate these values to the mean halo mass of the host halos. \cite{Sheth2001} extended the formalism of \cite{Mo1996} to account for the ellipsoidal collapse of dark matter halos. This gives the relation between bias and halo mass,
\begin{equation}
\label{eqn:biasmodel}
\begin{split}
b(M,z) = 1 + \frac{1}{\sqrt{a}\delta_{c}(z)}\bigg[ a\nu^{2}\sqrt{a}+0.5\sqrt{a}(a\nu^{2})^{1-c} \\
- \frac{ (a\nu^{2})^{c} }{ (a\nu^{2})^{c} + 0.5(1-c)(1-c/2) } \bigg],
\end{split}
\end{equation}
where $\nu=\delta_{c} / \sigma(M,z)$, $a=0.707$ and $c=0.6$. $\delta_{c}$ is the critical overdensity for the collapse of a homogeneous spherical perturbation, given by $\delta_{c}=0.15(12\pi)^{2/3}\Omega_{m}(z)^{0.0055}$, \citep{Navarro1997}. We describe the variance in the mass fluctuation of the density field for a mass scale $M$ as $\sigma(M)$
\begin{equation}
\sigma^{2}(M) = \frac{1}{ 2\pi^{2} }\int^{\infty}_{0}k^{2}P(k)w^{2}(kr)dk,
\end{equation}
where $P(k)$ is the matter power spectrum and 
\begin{equation}
w(kr)=\frac{ 3(kr\sin(kr)-\cos(kr)) }{(kr)^{3}},
\end{equation}
is the Fourier transform of a spherical tophat \citep{Peebles1980}. Radius $r$ is related to mass by
\begin{equation}
r=\Big( \frac{3M}{4\pi\rho_{0}} \Big)^{\frac{1}{3}},
\end{equation}
where $\rho_{0}$ is the mean density of the Universe at $z=0$, $\rho_{0}=2.78 \times 10^{11} \Omega_{m}^{0}h^{2}\mathrm{M_{\odot}}\mathrm{Mpc}^{-3}$. We calculate the rms mass fluctuation at a given redshift using the linear growth factor $D(z)$
\begin{equation}
\sigma(M,z) = \sigma(M)D(z).
\end{equation}
We show the bias values and associated halo masses in Table \ref{tab:absolute_redshift_table}. In Section \ref{sec:absmagred} we found little signal of a luminosity dependence of quasar clustering from our measurements of $r_{0}$. We compare the halo masses for different magnitude bins to re-examine those results. It is at higher redshift that we are best able to distinguish between different mass halos from their bias values as such we exclude the faintest luminosity bin as there was no data at higher redshifts.

The clustering of the remaining three magnitude bins is best described by halo masses of $6\pm8 \times 10^{12} h^{-1} \mathrm{M_{\odot}}$, $1.9\pm1.4 \times 10^{12} h^{-1} \mathrm{M_{\odot}}$ and $2.2\pm0.2 \times 10^{12} h^{-1} \mathrm{M_{\odot}}$ (rms error) for the $-24.5{<}M_{i}{<}-23.5$, $-25.5{<}M_{i}{<}-24.5$ and $-28.5{<}M_{i}{<}-25.5$ bins respectively.

We find that (excluding the high-z, low-M bin) the evolution of bias with redshift is well described by a mean halo mass of $M=2\pm1 \times 10^{12} h^{-1} \mathrm{M_{\odot}}$ (c.f. $M=3\pm5 \times 10^{12} h^{-1} \mathrm{M_{\odot}}$ including this bin). We show the model prediction for this halo mass in Figure \ref{fig:bias} as a solid line. Within the errors, our bias measurements are consistent with a single halo mass at all redshifts and luminosities. 

Our measurement of the evolution of b$(z)$ is slightly different than that of \citetalias{Croom2005}, the determination of halo mass has large errors. As such, our best-fit halo mass is lower than that of \citetalias{Croom2005} but remains consistent at the $1\sigma$ level.

\subsection{Comparison to XMM-COSMOS quasar clustering}
\label{sec:xray}
The semi-analytic model presented by \cite{Fanidakis2013} predict that  X-ray selected quasars inhabit higher mass halos than optically selected quasars. \cite{Fanidakis2013} present halo mass estimates from \cite{Allevato2011} and \cite{Krumpe2012} are presented as observational support to this model as these halo masses are higher (${\sim}10^{13}\mathrm{M_{\odot}}$) than estimates from wide area optical studies (${\sim}10^{12}\mathrm{M_{\odot}}$). In this section we briefly examine to whether this difference in halo mass estimates may be reconciled with the lack of dependence on optical luminosity found here. In particular, differences may occur due to differing analysis methods, and so we apply our method used for our optically selected samples to the  X-ray selected sample of \cite{Allevato2011}.  
  
\cite{Allevato2011} measured the correlation function for quasars in the XMM-COSMOS field \citep{Brusa2010} and found a clustering scale of $r_{0} = 7.08^{+0.30}_{-0.28} h^{-1}$ Mpc and $\gamma = 1.88^{+0.04}_{-0.06}$. We examine the sample of quasars used in their work and that find $g_{\mathrm{median}}{=}21.4$ (${\sim}0.1$ magnitudes fainter than the 2SLAQ sample) and their space density of quasars is ${\sim}90$deg$^{-2}$ which is similar to that reached by 2QDESp, see Table \ref{tab:summary}. Further, the redshift distribution of their  X-ray selected sources \citep[Figure 2; ][]{Allevato2011} is comparable to those of optically selected studies (see Figure \ref{fig:nz}). As we find no evidence for $r_{0}$ increasing with fainter magnitude, we believe their contradictory result worthy of further scrutiny.

Firstly, we note that an earlier clustering analysis of the XMM-COSMOS quasars \citep[${\sim}10\%$ fewer quasars than][]{Allevato2011} was performed by \cite{Gilli2009} who measured $r_{0} = 7.03^{+0.96}_{-0.89} \: h^{-1}$ Mpc with $\gamma{=}1.8$. We use the R.A.${-}$Dec. mixing approach of \cite{Gilli2009} to generate a random catalogue. However, instead of measuring $w(r_p)$ we measure the redshift correlation function, $\xi(s)$, for these data, assuming $\gamma=1.8$ as in Section \ref{sec:apparent} for the fit. \cite{Gilli2009} compared this method of random generation to modelling the angular distribution and found that it can underestimate the true correlation length. Applying the correction from \cite{Gilli2009} we find that the amplitude of clustering is described by $r_{0}{=}6.03^{+0.80}_{-1.00} \: h^{-1}$ Mpc. This is in agreement with the measurements of quasar clustering at $z{\approx}1.5$ found in this work. 

Both the $r_{0}$ measurements from \cite{Gilli2009} and \cite{Allevato2011} use the projected correlation function, $w(r_{p})$, as opposed to the redshift-space correlation function, $\xi(s)$, that we use. By remeasuring the correlation function we are able to compare directly to optical results.  As noted by other authors \citep{Mountrichas2012,Krumpe2012} this approach should provide a more robust comparison than comparing between different bias or halo mass models. 

We also note that our errors \citep[and those of][]{Gilli2009} assume Poisson statistics and still lead to a factor of $2{-}3\times$ larger errors on $r_{0}$ than the ${\approx}{\pm}0.3 \: h^{-1}$Mpc quoted by \cite{Allevato2011}; it is not clear why this is the case. If the statistical errors on the XMM-COSMOS results are as large as found by \cite{Gilli2009} and ourselves then we conclude that these data contain no significant evidence for luminosity dependent clustering e.g. compared to their brighter counterparts in Figure \ref{fig:foo} (see also discussion in next section).

\subsection{Baryonic Acoustic Oscillations}
\begin{figure}
\centering
\includegraphics[width=\linewidth]{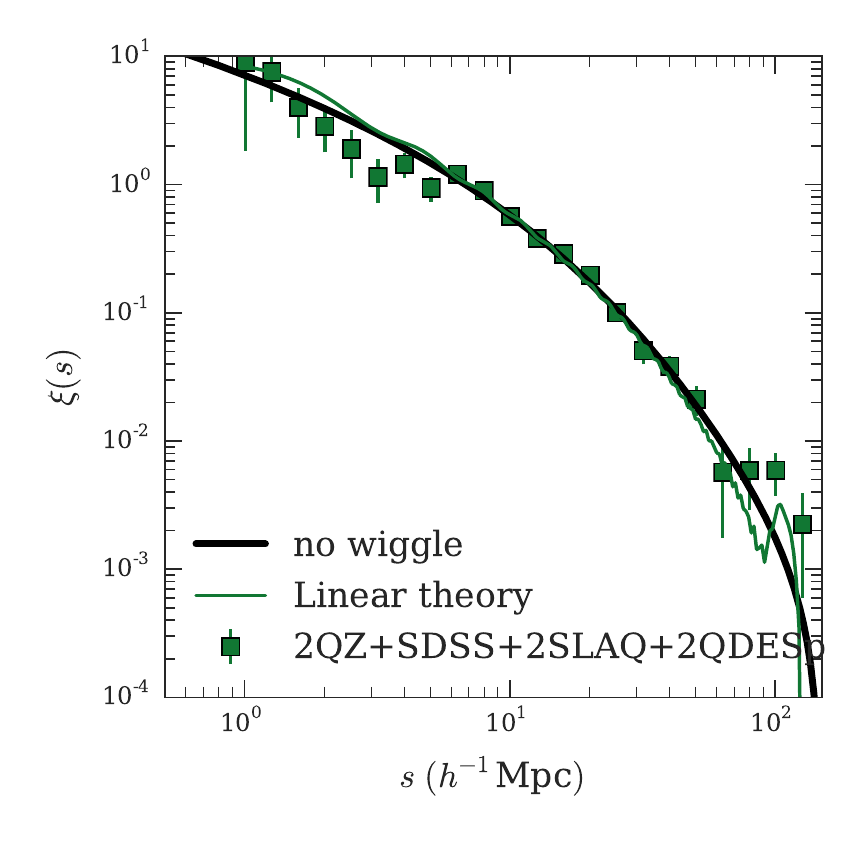}
\caption{At each $s$ bin we combine the values of the correlation function $\xi(s)$ for each of the four surveys using the error weighted mean. This combined sample consists of $N_{QSO}=70 \: 940$ with mean redshift $\bar{z}=1.49$. We fit our model form Section \protect{\ref{sec:model}} to the data and find a best-fit value for $r_{0} = 6.10{\pm}0.10 \: h^{-1} Mpc$ with $\chi^{2},$df.$=15.6,10$ where we fit in the range $5{<} s (h^{-1} Mpc) {<}55$ (shown as the solid line). We also include the prediction of linear theory from {\tt CAMB} and the `no wiggle' model of \protect{\citet{Eisenstein1998}} with both normalised to our correlation function amplitude between $5{<}s \: (h^{-1} \mathrm{Mpc}){<}55$.}
\label{fig:combined}
\end{figure}

Here we extend our analysis of the combined quasar sample to larger scales. In Figure \ref{fig:combined} we show the result of combining the
four correlation functions from each of the four surveys, weighting inversely according to the square of the errors at each separation
(section \ref{sec:apparent}). We measure $r_{0}{=}6.10{\pm}0.15 h^{-1}$Mpc for a sample containing $70 940$ quasars with ${<}z{>}=1.49$. Combining
these surveys gives an effective volume of ${\approx}0.6 \: h^{-3}$Gpc$^{3}$, larger than the original SDSS LRG survey of \cite{Eisenstein2005} (${\approx}0.55 \: h^{-3}$Gpc$^{3}$ or the 2dFGRS survey of \citet{Cole2005} (${\approx}0.1 \: h^{-3}$Gpc$^{3}$) where BAO were detected. We use {\tt \sc CAMB} to predict the $\Lambda$CDM correlation function and scale this model to agree with the measured $\xi(s)$ at intermediate scales, $5{<}s{<}55 \: h^{-1}$Mpc (see Figure \ref{fig:combined}). Comparing the model to the data $\xi(s)$ at larger scales, $60{<}s{<}200 \: h^{-1}$Mpc, we find that the model with the BAO feature is fit with $\chi^{2},$df.=5.5,4 and p-value=0.23 compared to $\chi^{2},$df.$=6.1,4$ and p-value${=}0.19$ for a similar model without BAO. Although the model with the BAO feature fits better, the reduction in $\chi^2$ is not significant ($\Delta \chi^2\approx0.6$) and so it is not possible to claim that the BAO feature is detected in this combined quasar survey.

We consider possible explanations for this lack of detection. Firstly, the statistical errors are still relatively large, larger still once the off-diagonal covariance matrix elements are considered, motivating the need for bigger samples with larger effective volumes at the BAO scale. However, we have argued above that the effective volume should already be large enough for the detection of this feature. Secondly, it does not appear that the ${\pm}750$km s$^{-1}$ quasar redshift error plus intrinsic velocity dispersion affect the detectability of the BAO peak, as evidenced by convolving the $\Lambda$CDM model with Gaussians of this width. Our $9\%$ fraction of misidentified quasar redshifts will reduce the BAO signal and the small scale signal in proportion and so this effect has already been accounted for in Figure \ref{fig:combined} by our procedure of scaling the model to the observed small-scale $\xi(s)$. Thus it remains unclear why the BAO peak is undetected in these data.
%
%

\section{Discussion}
\label{sec:dis}
We have analysed quasar clustering using surveys covering a wide range of fluxes and luminosities. We have found that there is little evidence
for an increase of clustering amplitude with luminosity within the optical surveys at fixed redshift. Even including the XMM-COSMOS survey,
we still find no evidence for the dependence of the clustering scale on luminosity. Following \citetalias{Shanks2011} we assume a halo-black hole mass relation of the form $M_{BH}{\sim} M_{Halo}^{1.82}$ \citep{Ferrarese2002} and the bias $b{\sim} M_{Halo}^{0.2}$ \citep{Martini2001} together with a fixed Eddington ratio we expect the approximate relation $r_{0} \propto b \propto M_{BH}^{0.1} \propto L^{0.1}$. Given the factor of ${\approx}10$ increase in luminosity between the SDSS and 2SLAQ samples, a factor of ${\approx}1.25$ increase in $r_{0}$ is predicted, corresponding to $r_{0}$ increasing from $6.20$ (2SLAQ) to $7.75\:h^{-1}$Mpc, significantly (${\approx}4\sigma$) higher than the observed value from SDSS. Thus, the observed luminosity dependence of the clustering amplitude is about a fifth of what is predicted on the basis of this simple model. This is confirmed by the formal $\chi^2$ fits of the $L^{0.1}$ model in Figure \ref{fig:foo}. Excluding XMM-COSMOS, we find $\chi^{2},$df.=7.8,3 and $p$-value=0.05 for the $L^{0.1}$ model compared to $\chi^{2},$df.=3.8,3 and $p$-value=0.23 for the $L{-}$independent case. Including XMM-COSMOS, the same preference for the $L$ independent model is shown, although the $L^{0.1}$ model is slightly less rejected with $p$-value=0.10. But if the XMM-COSMOS $r_{0}$ was closer to its corrected value of $r_{0}{=}6.0\:h^{-1}$Mpc, rather than the $r_{0}{=}5.2\:h^{-1}$Mpc we have assumed here, then the level of rejection of the $L^{0.1}$ model would again be increased. We conclude that any dependence of clustering amplitude with luminosity is smaller than expected from a simple halo model.  

When we then sub-divided the combined SDSS, 2QZ, 2SLAQ and 2QDESp surveys by absolute magnitude and redshift to increase the dynamic range in luminosity, we again found no evidence for luminosity dependent clustering at fixed redshift. However, we note that we do have significant evidence for the dependence of $r_{0}$ on redshift. We introduced a new bias model for $b(z)$ (equation \ref{eqn:empbias}), superseding that of \citetalias{Croom2005}. We find that our model for the evolution of bias with redshift is consistent with  the higher $r_{0}=7.25{\pm}0.10\:h^{-1}$Mpc measured by \cite{Eftekharzadeh2015} in the BOSS quasar survey at $z{\approx}2.4$. 

The 2QZ results of \citetalias{Croom2005} suggest that a fixed halo mass of ${\sim}10^{12.5}$M$_\odot$ fits the $z$ dependence of quasar clustering. Here we have confirmed the results of \citetalias{Shanks2011} that at approximately fixed redshift the clustering amplitude is approximately constant with luminosity. The apparent luminosity independence would suggest that the halo mass and hence black hole mass was virtually constant as a function of both luminosity and redshift. 

If all quasars have the same black hole mass over a wide range of luminosity then there is a contradiction with all quasars radiating at a fixed fraction of Eddington as suggested from reverberation studies of nearby quasars \citep{Peterson2004}. To reconcile these two observations requires that the BH${-}$halo mass relation is broken. There is some evidence for this from the work of \cite{Kormendy2011} who found that the BH mass was more related to bulge mass than halo mass. In this view quasars would find themselves in similar sized haloes but with black hole masses  more related to their luminosity.

A weak clustering dependence on luminosity is expected in `flickering' models where the duty cycle for AGN activity is short and the quasar luminosity is highly variable \citep{Lidz2006}. The luminosity of a quasar may also be taken as implying a lower limit to its black-hole mass via the Eddington limit. Thus low luminosity quasars must be accreting at a highly sub-Eddington rate if the halo-mass BH mass relation is to be preserved, since they have similar halo masses to their brighter counterparts. This means that quasars are preferentially seen in bright phases. \citetalias{Shanks2011} again noted that this model contradicts the established correlation from reverberation mapping between black hole mass and luminosity \citep{Peterson2004} but otherwise fits the clustering data (by design).
 
In Section \ref{sec:bias} we estimate the halo bias for optically selected quasars between $0.3{<}z{<}2.9$ and $-28.5{<}M_{i(z=2)}{<}-23.5$. In agreement with earlier works \citepalias{Croom2005,daAngela2008,Ross2009,Shanks2011} we find a characteristic halo mass of $M_{\mathrm{Halo}}{=}2{\pm}1{\times}10^{12} h^{-1} \mathrm{M_{\odot}}$. Recent measurements of quasar clustering from  X-ray surveys \citep{Gilli2009,Allevato2011,Krumpe2012} have estimated significantly higher halo masses (${\sim}10^{13}\mathrm{M_{\odot}}$) than the above optically selected samples. 

Semi-analytical models of AGN \citep{Fanidakis2013} have suggested that this is a physical result caused by a difference in AGN fuelling modes between optically selected and  X-ray selected samples. However, given the susceptibility of soft  X-ray selection (${\approx}0.1-2keV$) to intrinsic obscuration we would expect these two selections to sample the same population of AGN. This is supported by the similar space density and redshift distribution of unobscured  X-ray AGN and optically selected quasars \citep[see][]{Allevato2011}.

Indeed, both \cite{Krumpe2012} and \cite{Allevato2011} explicitly compare the clustering of optically selected quasars with unobscured  X-ray AGN in their two samples. In both papers these authors find that the clustering of both populations (at any redshift) may be described by the same halo mass. Contrary to the claim of \cite{Allevato2011} we find consistent clustering of  X-ray and optically selected samples. As such, we see little evidence for the higher halo masses in these studies (c.f. optical studies) that would support the suggestion of \cite{Fanidakis2013} that the two populations are driven by different accretion modes.  The analysis by \cite{Mountrichas2013} suggests that higher X-ray AGN masses are in fact driven by X-ray AGN from groups. After excluding these AGN, \cite{Mountrichas2013} find the clustering of X-ray selected AGN is described by a halo mass $5^{+4}_{{-}3}{\times}10^{12} \: h^{-1} \mathrm{M_{\odot}}$, consistent with the clustering results presented here.

\cite{Krumpe2012} discuss the impact of HOD vs. power-law models for estimating bias from the correlation function. We agree that this may contribute to the differences in halo mass estimates. We further note that the deepest  X-ray samples come from small areas on the sky and suffer from poorer statistics and greater susceptibility to sample variance than the larger area optical studies. This discrepancy could be well addressed by a large sample of deep  X-ray selected AGN. Ongoing surveys such as eBOSS and the upcoming eROSITA survey have the opportunity to provide a homogeneous dataset of quasars up to $z{\lesssim}2.2$. This may allow us to determine which physical processes drive accretion at different redshifts and how these processes interact to result in quasar clustering appearing largely independent of optical luminosity.

\section{Conclusions}
We have characterised a new quasar selection for quasars at intermediate redshifts $0.8{\lesssim}z{\lesssim}2.5$ and we demonstrate that the WISE
All-Sky data release is complete for quasars in the redshift range (with $g{<}20.5$). To account for photometric incompleteness for quasars fainter
than this limit, to $g{\approx}22.5$, requires traditional optical selection methods. 

The 2QDES pilot survey has shown that a high density quasar survey is viable with the photometry from VST ATLAS and WISE. In fact the 2QDESp
survey with $4\%$ of the area of SDSS has $20\%$ more effective volume to detect the BAO peak due to its ${\approx}8{\times}$ higher quasar density. But even with $10 000$ quasars from 2QDESp combined with those from SDSS, 2QZ and 2SLAQ, we still lack a large enough effective volume
to measure the BAO peak in the two-point correlation function although we gain some advantage in the precision of the clustering scale length,
$r_{0}$.

Direct comparison between the quasar correlation functions of SDSS, 2QZ, 2QDESp and 2SLAQ surveys, that range over an order of magnitude in
quasar luminosity, show the same hint of higher $r_{0}$ at higher (SDSS) luminosities that was seen by \cite{Shen2009} and
\citetalias{Shanks2011}. However, the errors are such that a luminosity independent $r_{0}$ cannot be rejected by these data.

We combine the clustering measurements from 2QZ, 2SLAQ, SDSS and 2QDESp in the $M-z$ plane to search further for luminosity and redshift
dependence. Contrary to the above hint, we find some tentative evidence here that \textit{fainter} quasars may be  more \textit{strongly}
clustered than brighter quasars at fixed (high) redshift ($z{>}1.5$), albeit weakly detected. But overall the results remain consistent 
with a fixed quasar $r_{0}$ at fixed redshift, independent of luminosity.

We measure a  significant redshift dependence of quasar clustering and see that this dependence explains the higher $r_{0}$
measurements from \cite{Eftekharzadeh2015}. Comparison of the redshift dependence of quasar clustering to the halo model shows that our data
(and that of \cite{Eftekharzadeh2015}) is consistent with quasars inhabiting $2 \times 10^{12} \mathrm{M_{\odot}}$ halos irrespective of redshift
or quasar luminosity. These results are usually explained by a `flickering' quasar model with a short duty cycle where quasars over a
wide range of luminosities have similar halo, and hence black hole, masses. However, such models are inconsistent with the strong
correlation between black hole and luminosity found from reverberation mapping. \cite{Shanks2011} indicated that the quasar clustering and
reverberation mapping results might only be reconciled by breaking the black hole mass-halo mass correlation, as suggested by \cite{Kormendy2011}.

We also found similar clustering scale lengths ($r_{0}\approx6\:h^{-1}$Mpc for quasars in the XMM-COSMOS field, with little evidence that such quasars show a higher clustering amplitude than their more luminous, optically selected counterparts, as previously reported. This means that there is less evidence for higher halo masses at low redshift for AGN accreting in the hot halo mode, as suggested by \cite{Fanidakis2013}.

\section{Acknowledgments}
\label{sec:ack}
The VST-ATLAS survey is based on observations made with ESO telescopes at the La Silla Paranal Observatory under programme ID 177.A-3011. We are indebted to the CASU for reducing the ATLAS images and producing catalogues. This publication makes use of data products from the Wide-field Infrared Survey Explorer, which is a joint project of the University of California, Los Angeles, and the Jet Propulsion Laboratory/California Institute of Technology, funded by the National Aeronautics and Space Administration. The 2QDES pilot survey was based on observations made with the AAT and we would like to thank the staff of the Australian Astronomical Observatory for their contribution to this work. We thank Nicholas Ross for useful discussions on the quasar selection and the anonymous referee for improving the quality of this manuscript. This work makes use of the {\tt runz } redshifting code developed by Will Sutherland, Will Saunders, Russell Cannon and Scott Croom. BC, TS, SF and DP are incredibly grateful for the efforts of the NSW Rural Fire Service who worked to ensure our safe evacuation of the site during the 2013 bushfire. This work was supported by the Science and Technology Facilities Council [ST/K501979/1,ST/L00075X/1]. 
\section{References}
\bibliographystyle{mnras} 
\bibliography{quasar_clustering_v11} 
\newpage
\appendix
\section{Details of observations}
\begin{table*}
\begin{tabular*}{\textwidth}{c @{\extracolsep{\fill}} llllllllllll}
\hline
\hline
Field & Date & $N_{Stars}$ & $N_{Q}$ & $N_{Q}{\leq}20.5$ & $\frac{S}{N}$& QSO$_{lim}$ & $u_{lim}$ & Algorithm & Repeat & Comment \\
\hline
\scriptsize hh:mm$\pm$dd:mm& &   deg$^{-2}$          &    deg$^{-2}$       & deg$^{-2}$                    &         & [AB] & [AB]     &           &        &      \\
\hline
0231$-$3002 & 2011:12:20 & 1881 & 64.33 & 28.03 & 2.55 & 21.72 & 22.13 & $ugri${\tt XDQSO} & 0 & \\
1046$-$0704 & 2012:04:28 & 3122 & 26.11 & 14.65 & 1.25 & 21.14 & 21.93 & $ugri${\tt XDQSO} & 0 & \\
1236$-$0704 & 2012:04:28 & 3427 & 37.90 & 20.06 & 2.76 & 21.21 & 22.13 & $ugri${\tt XDQSO} & 0 & \\
1301$-$0704 & 2012:04:28 & 3569 & 46.50 & 20.38 & 2.29 & 21.48 & 22.16 & $ugri${\tt XDQSO} & 0 & \\
1004$-$0704 & 2012:04:28 & 3595 & 47.13 & 19.11 & 3.00 & 21.50 & 22.01 & $ugri${\tt XDQSO} & 0 & \\
1444$-$0704 & 2012:04:28 & 5422 & 27.39 & 15.61 & 0.93 & 21.72 & 22.10 & $ugri${\tt XDQSO} & 0 & \\
2122+0000 & 2012:04:28 & 8850 & 39.17 & 17.20 & 0.96 &   21.99 & 21.87\textdagger & $ugri${\tt XDQSO} & 0 & \\
1219$-$0704 & 2012:04:29 & 3346 & 49.36 & 23.89 & 2.38 & 21.57 & 22.02 & $ugri${\tt XDQSO} & 0 & \\
1244$-$0704 & 2012:04:29 & 3483 & 45.22 & 22.29 & 2.27 & 21.37 & 22.22 & $ugri${\tt XDQSO} & 0 & \\
1309$-$0704 & 2012:04:29 & 3563 & 50.32 & 19.75 & 2.48 & 21.51 & 22.10 & $ugri${\tt XDQSO} & 0 & \\
1453$-$0704 & 2012:04:29 & 5451 & 37.90 & 23.25 & 2.29 & 21.52 & 22.05 & $ugri${\tt XDQSO} & 0 & \\
0300+0000   & 2012:11:04 & 2221 & 45.22 & 22.29 & 1.65 & 22.07 & 21.87\textdagger & $ugri${\tt XDQSO} & 0 & \\
2342$-$3102 & 2012:11:04 & 2412 & 61.47 & 24.52 & 4.68 & 21.55 & 22.11 & $ugri${\tt XDQSO} & 0 & \\
2251$-$3102 & 2012:11:04 & 3200 & 60.19 & 22.61 & 3.74 & 21.57 & 22.08 & $ugri${\tt XDQSO} & 0 & \\
0346$-$2604 & 2013:01:11 & 2264 & 71.97 & 24.20 & 4.48 & 21.88 & 22.04 & $ugri${\tt XDQSO}W1W2 & 0 & \\ 
0356$-$2903 & 2013:01:11 & 2458 & 72.61 & 34.08 & 4.29 & 21.77 & 22.02 & $ugri${\tt XDQSO}W1W2 & 0 & \\ 
1023$-$0903 & 2013:01:11 & 3668 & 50.64 & 21.97 & 4.97 & 21.76 & 21.81 & $ugri${\tt XDQSO}W1W2 & 0 & Twilight \\ 
1004$-$0803 & 2013:01:11 & 3771 & 59.55 & 21.34 & 3.48 & 21.61 & 21.92 & $ugri${\tt XDQSO}W1W2 & 0 & \\ 
0356$-$2604 & 2013:01:12 & 2320 & 64.65 & 29.62 & 2.50 & 21.73 & 22.08 & $ugri${\tt XDQSO}W1W2 & 0 & \\ 
1055$-$0903 & 2013:01:12 & 3478 & 55.10 & 23.57 & 2.66 & 21.66 & 21.88 & $ugri${\tt XDQSO}W1W2 & 0 & Twilight \\ 
1016$-$0903 & 2013:01:12 & 3818 & 50.00 & 25.48 & 2.93 & 21.62 & 21.83 & $ugri${\tt XDQSO}W1W2 & 0 & \\ 
0352$-$2804 & 2013:02:16 & 2518 & 67.52 & 30.57 & 2.78 & 21.62 & 21.88 & $ugri${\tt XDQSO}W1W2 & 0 & Moon \\ 
1029$-$0706 & 2013:02:16 & 3192 & 50.32 & 25.16 & 2.47 & 21.38 & 22.00 & $ugri${\tt XDQSO}W1W2 & 0 & \\ 
1218$-$0704 & 2013:02:16 & 3336 & 52.23 & 29.94 & 2.61 & 21.42 & 22.01 & $ugri${\tt XDQSO}W1W2 & 0 & \\ 
1227$-$0704 & 2013:02:16 & 3363 & 58.92 & 30.89 & 2.45 & 21.51 & 22.07 & $ugri${\tt XDQSO}W1W2 & 0 & \\ 
1021$-$0706 & 2013:02:16 & 3435 & 50.00 & 23.57 & 2.26 & 21.59 & 22.00 & $ugri${\tt XDQSO}W1W2 & 0 & \\ 
1012$-$0706 & 2013:02:16 & 3454 & 59.24 & 31.21 & 3.16 & 21.68 & 22.02 & $ugri${\tt XDQSO}W1W2 & 0 & \\ 
0330$-$3002 & 2013:02:17 & 2176 & 73.25 & 32.17 & 3.47 & 22.04 & 22.04 & $ugri${\tt XDQSO}W1W2 & 0 & Moon \\ 
1038$-$0706 & 2013:02:17 & 2931 & 57.64 & 25.48 & 2.97 & 21.63 & 21.98 & $ugri${\tt XDQSO}W1W2 & 0 & Moon \\ 
1046$-$0706 & 2013:02:17 & 3148 & 68.79 & 35.67 & 3.64 & 21.49 & 21.92 & $ugri${\tt XDQSO}W1W2 & 0 & \\ 
1055$-$0706 & 2013:02:17 & 3219 & 60.19 & 30.25 & 2.83 & 21.47 & 21.94 & $ugri${\tt XDQSO}W1W2 & 0 & \\ 
1235$-$0704 & 2013:02:17 & 3432 & 50.32 & 25.48 & 3.24 & 21.81 & 22.13 & $ugri${\tt XDQSO}W1W2 & 0 & \\ 
1252$-$0704 & 2013:02:17 & 3525 & 55.10 & 24.20 & 2.75 & 21.62 & 22.32 & $ugri${\tt XDQSO}W1W2 & 0 & \\ 
1318$-$0704 & 2013:02:17 & 3642 & 64.01 & 30.89 & 2.51 & 21.68 & 22.23 & $ugri${\tt XDQSO}W1W2 & 0 & \\ 
0339$-$3002 & 2013:02:18 & 2295 & 59.87 & 34.39 & 3.88 & 21.58 & 22.17 & $ugri${\tt XDQSO}W1W2 & 0 & Moon $+$ Artefact \\
1046$-$0906 & 2013:02:18 & 3336 & 51.59 & 23.89 & 2.89 & 21.84 & 21.85 & $ugri${\tt XDQSO}W1W2 & 0 & Artefact \\ 
1038$-$0906 & 2013:02:18 & 3572 & 48.09 & 28.66 & 3.03 & 21.88 & 21.90 & $ugri${\tt XDQSO}W1W2 & 0 & Moon $+$ Artefact \\
1434$-$0704 & 2013:02:18 & 5158 & 50.00 & 21.02 & 2.83 & 21.61 & 22.09 & $ugri${\tt XDQSO}W1W2 & 0 & Artefact\\ 
1442$-$0704 & 2013:02:18 & 5408 & 49.36 & 21.97 & 2.92 & 21.82 & 22.10 & $ugri${\tt XDQSO}W1W2 & 0 & Artefact\\ 
1508$-$0704 & 2013:02:18 & 6623 & 49.68 & 22.61 & 3.25 & 21.67 & 22.07 & $ugri${\tt XDQSO}W1W2 & 0 & Artefact\\ 
0349$-$3002 & 2013:02:19 & 2392 & 59.87 & 33.44 & 5.74 & 21.68 & 22.13 & $ugri${\tt XDQSO}W1W2 & 0 & Moon $+$ Artefact \\ 
1103$-$0906 & 2013:02:19 & 3253 & 54.46 & 23.89 & 4.14 & 21.85 & 21.86 & $ugri${\tt XDQSO}W1W2 & 0 & Moon $+$ Artefact \\ 
1103$-$0905 & 2013:02:19 & 3275 & 23.25 & 10.83 & 3.32 & 21.85 & 21.86 & $ugri${\tt XDQSO}W1W2 & 0 & Artefact, repeated in error \\ 
1451$-$0704 & 2013:02:19 & 5344 & 59.55 & 27.07 & 3.18 & 21.63 & 22.05 & $ugri${\tt XDQSO}W1W2 & 0 & Artefact \\ 
1459$-$0704 & 2013:02:19 & 6186 & 33.44 & 13.38 & 3.25 & 21.82 & 22.06 & $ugri${\tt XDQSO} & 0 & Artefact \\ 
1516$-$0704 & 2013:02:19 & 7287 & 45.22 & 19.11 & 2.59 & 21.63 & 22.05 & $ugri${\tt XDQSO}W1W2 & 0 & Twilight $+$ Artefact \\ 
2307$-$2601 & 2013:07:28 & 2981 &  9.87 &  2.23 & 1.17 & 21.60 & 22.07 & $ugri${\tt XDQSO}W1W2 & 1 & \\ 
2307$-$2601 & 2013:07:28 & 2982 & 55.10 & 26.75 & 2.35 & 21.60 & 22.07 & $ugri${\tt XDQSO}W1W2 & 0 & \\ 
1442$-$1500 & 2013:07:28 & 6865 & 57.01 & 19.75 & 2.41 & 21.93 & 21.80 & $ugri${\tt XDQSO}W1W2 & 0 & \\ 
1442$-$1500 & 2013:07:28 & 6868 & 11.78 &  1.91 & 1.67 & 21.93 & 21.80 & $ugri${\tt XDQSO}W1W2 & 1 & \\ 
1500$-$1500 & 2013:07:28 & 7987 & 52.23 & 24.20 & 3.05 & 21.75 & 21.96 & $ugri${\tt XDQSO}W1W2 & 0 & \\ 
1500$-$1500 & 2013:07:28 & 7990 &  8.60 &  1.59 & 1.92 & 21.75 & 21.96 & $ugri${\tt XDQSO}W1W2 & 1 & \\ 
2316$-$2601 & 2013:07:29 & 2957 & 69.11 & 33.76 & 2.63 & 21.75 & 22.16 & $ugri${\tt XDQSO}W1W2 & 0 & Cloud in Field \\ 
2133$-$2807 & 2013:07:29 & 5579 & 17.52 &  2.23 & 1.63 & 21.69 & 21.98 & $ugri${\tt XDQSO}W1W2 & 1 & Cloud in Field \\ 
2133$-$2807 & 2013:07:29 & 5581 & 45.54 & 26.11 & 1.66 & 21.69 & 21.98 & $ugri${\tt XDQSO}W1W2 & 0 & Cloud in Field \\ 
1451$-$1500 & 2013:07:29 & 7389 & 44.90 & 23.57 & 6.32 & 21.79 & 21.86 & $ugri${\tt XDQSO}W1W2 & 0 & \\ 
1451$-$1500 & 2013:07:29 & 7390 & 13.38 &  1.91 & 2.32 & 21.75 & 21.86 & $ugri${\tt XDQSO}W1W2 & 1 & \\ 
1508$-$1500 & 2013:07:29 & 8390 & 38.53 & 15.29 & 1.57 & 21.78 & 21.93 & $ugri${\tt XDQSO}W1W2 & 0 & Cloud in Field \\ 
2258$-$2807 & 2013:07:30 & 2962 & 19.43 &  2.23 & 2.11 & 21.72 & 21.91 & $ugri${\tt XDQSO}W1W2 & 1 & \\ 
2258$-$2807 & 2013:07:30 & 2964 & 60.19 & 30.25 & 2.03 & 21.84 & 21.91 & $ugri${\tt XDQSO}W1W2 & 0 & \\ 
2133$-$2601 & 2013:07:30 & 5443 & 16.24 &  1.59 & 1.59 & 21.69 & 22.06 & $ugri${\tt XDQSO}W1W2 & 1 & \\ 
2133$-$2601 & 2013:07:30 & 5443 & 66.56 & 28.98 & 3.02 & 21.69 & 22.06 & $ugri${\tt XDQSO}W1W2 & 0 & \\ 
1508$-$1500 & 2013:07:30 & 8391 & 19.11 &  2.55 & 1.57 & 21.89 & 21.93 & $ugri${\tt XDQSO}W1W2 & 1 & \\ 
1526$-$1500 & 2013:07:30 & 9992 & 49.36 & 19.75 & 2.32 & 21.78 & 21.79 & $ugri${\tt XDQSO}W1W2 & 0 & \\ 
2316$-$2601 & 2013:07:31 & 2957 & 21.97 &  4.14 & 1.49 & 21.75 & 22.16 & $ugri${\tt XDQSO}W1W2 & 1 & \\ 
2215+0014   & 2013:07:31 & 4426 & 50.00 & 20.06 & 3.08 & 21.84 & 21.87\textdagger & $ugri${\tt XDQSO}W1W2 & 0 & \\ 
\hline
\end{tabular*}
\end{table*}
\begin{table*}
	\begin{tabular*}{\textwidth}{c @{\extracolsep{\fill}} llllllllllll}
		\hline
		\hline
		Field & Date & $N_{Stars}$ & $N_{Q}$ & $N_{Q}{\leq}20.5$ & $\frac{S}{N}$& QSO$_{lim}$ & $u_{lim}$ & Algorithm & Repeat & Comment \\
		\hline
		\scriptsize hh:mm$\pm$dd:mm& &   deg$^{-2}$          &    deg$^{-2}$       & deg$^{-2}$                    &         & [AB]     &[AB]     &           &        &      \\
		\hline
2152$-$2601 & 2013:07:31 & 4899 & 61.78 & 28.98 & 2.40 & 21.79 & 22.06 &  $ugri${\tt XDQSO}W1W2 & 0 & \\ 
2152$-$2601 & 2013:07:31 & 4901 & 15.92 &  0.96 & 1.36 & 21.79 & 22.06 &  $ugri${\tt XDQSO}W1W2 & 1 & \\ 
1447$-$1858 & 2013:07:31 & 8648 &  5.41 &  1.59 & 1.87 & 21.72 & 21.91 &  $ugri${\tt XDQSO}W1W2 & 1 & \\ 
1447$-$1858 & 2013:07:31 & 8650 & 53.82 & 20.38 & 3.81 & 21.72 & 21.91 &  $ugri${\tt XDQSO}W1W2 & 0 & \\ 
1526$-$1500 & 2013:07:31 & 9993 &  7.64 &  2.23 & 1.53 & 21.78 & 21.79 &  $ugri${\tt XDQSO}W1W2 & 1 & \\   
\hline
\label{tab:summary}
\end{tabular*}
\caption{$N_{Stars}$ are the number of point sources brighter than $g=21.5$, $N_{Q}$ is the number of spectroscopically confirmed quasars deg$^{-2}$ in that field, $\frac{S}{N}$ is the mean signal-to-noise per pixel. $N_{Q}{\leq}20.5$ are the number of spectroscopically confirmed quasars deg$^{-2}$ in a field brighter than $g=20.5$. QSO$_{lim}$ is the limit in g that contains 90\% of our quasar sample in that field. $u_{lim}$ is the average 5$\sigma$ limit in u for the stacked images that make up a 2dF field.The algorithm specifies whether we select quasars using the {\tt XDQSO} algorithm and optical colour selection alone or whether we used WISE photometry as well. Repeated fields are fields that were observed for a second time with new fibre allocations. \textdagger $5\sigma$ $u$-band limits for SDSS imaging are the characteristic SDSS limits as shown in \protect \cite{Shanks2014}.}
\end{table*}
\section{Covariance Matrix}
\label{appendix:matrix}
\begin{figure}
\includegraphics[width=\linewidth]{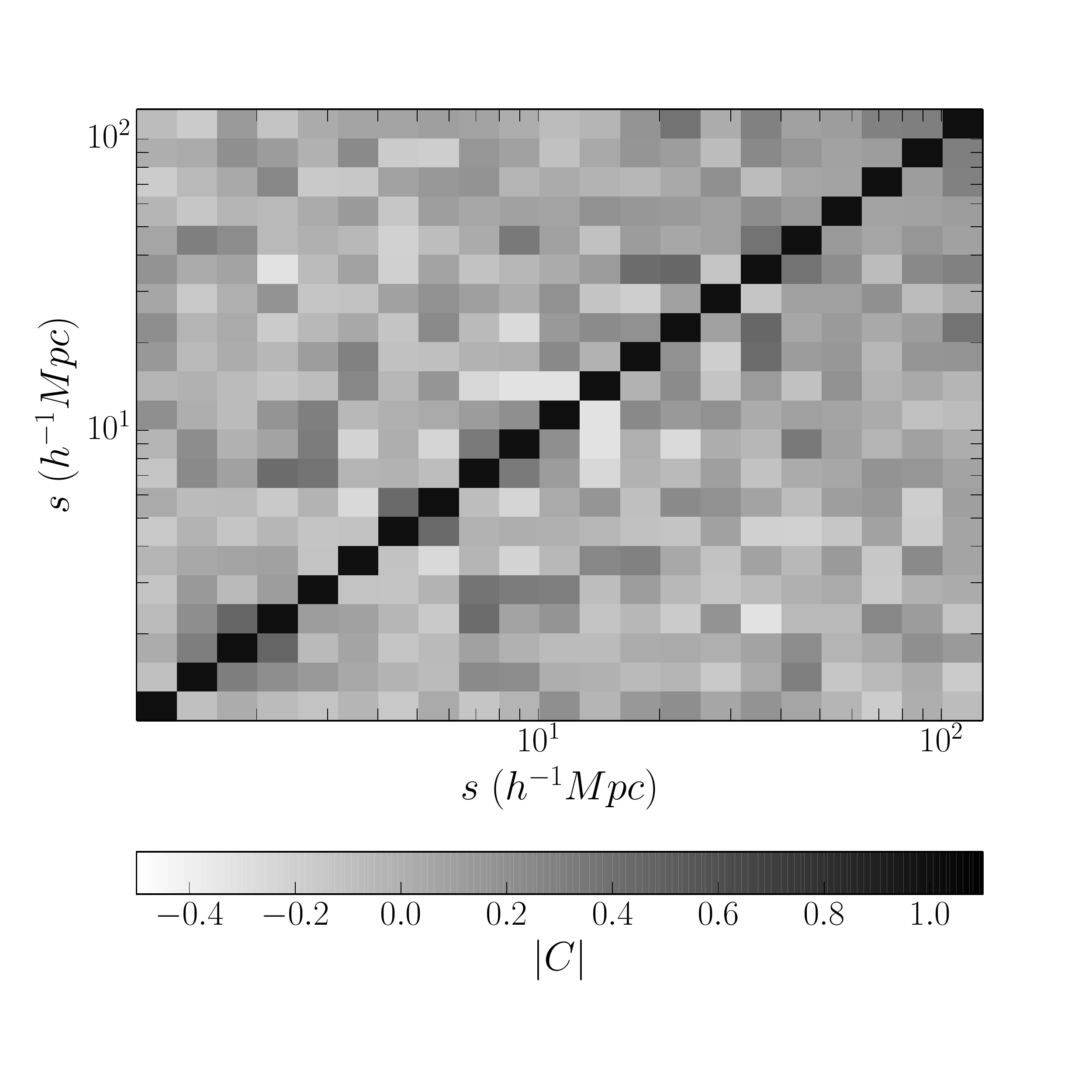}
\caption{The normalised covariance matrix (see Equation  \ref{eqn:normalised_covariance}) for $\xi(s)$ with jackknife errors calculated from dividing our sample into the separate 2dF pointings.}
\label{fig:covariance}
\end{figure}

We calculate the covariance matrix for our full sample, described in Section \ref{sec:qsocat}. Using a similar approach of \citetalias{Ross2009}, we calculate the inverse-variance weighted covaraince matrix, $C_{ij}$ by
\begin{multline}
C_{ij}=\sum_{L=1}^{N}\sqrt{\frac{DR_{L}(s_{i})}{DR(s_{i})}}[\xi_{L}(s_{i})-\xi_{total}(s_{i})] \times \\
\sqrt{\frac{DR_{L}(s_{j})}{DR(s_{j})}}[\xi_{L}(s_{j})-\xi_{total}(s_{j})]
\end{multline}
where $DR$ denotes the number of quasar-random pairs remaining when we \emph{exclude} subregion $L$ from the analysis. We recalculate $\xi_{L}$ (see Equation \ref{eqn:ls}) for the remaining sample, after excluding the specified region, $L$. In Figure \ref{fig:covariance} we present the covariance matrix for our sample. We normalise the matrix such that 
\begin{equation}
|C| = \frac{C_{ij}}{\sigma_{i}\sigma_{j}}
\label{eqn:normalised_covariance}
\end{equation}
\section{Observation pointings}
\label{appendix:pointings}
\onecolumn
\begin{figure}
	\includegraphics[width=\linewidth]{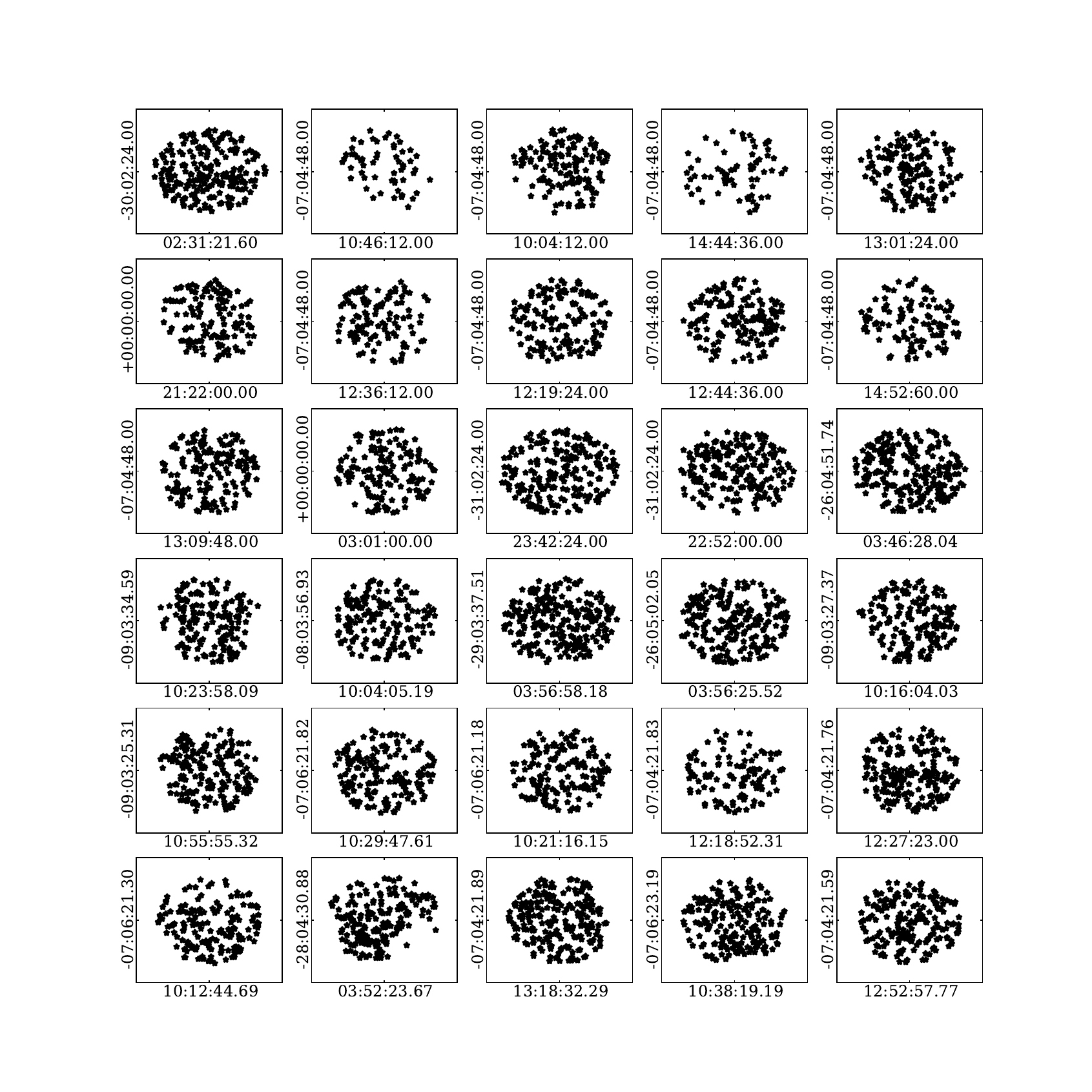}
	\caption{We show the quasar sample obtained from each 2dF pointing. The field at $11{:}03{:}49.48 -09{:}05{:}44.37$ was repeated in error. This provided us with duplicate redshifts for the same quasars. These were analysed to provide redshift error estimates found in Section \ref{sec:redshifterror}.}
\end{figure}
\begin{figure}
	\includegraphics[width=\linewidth]{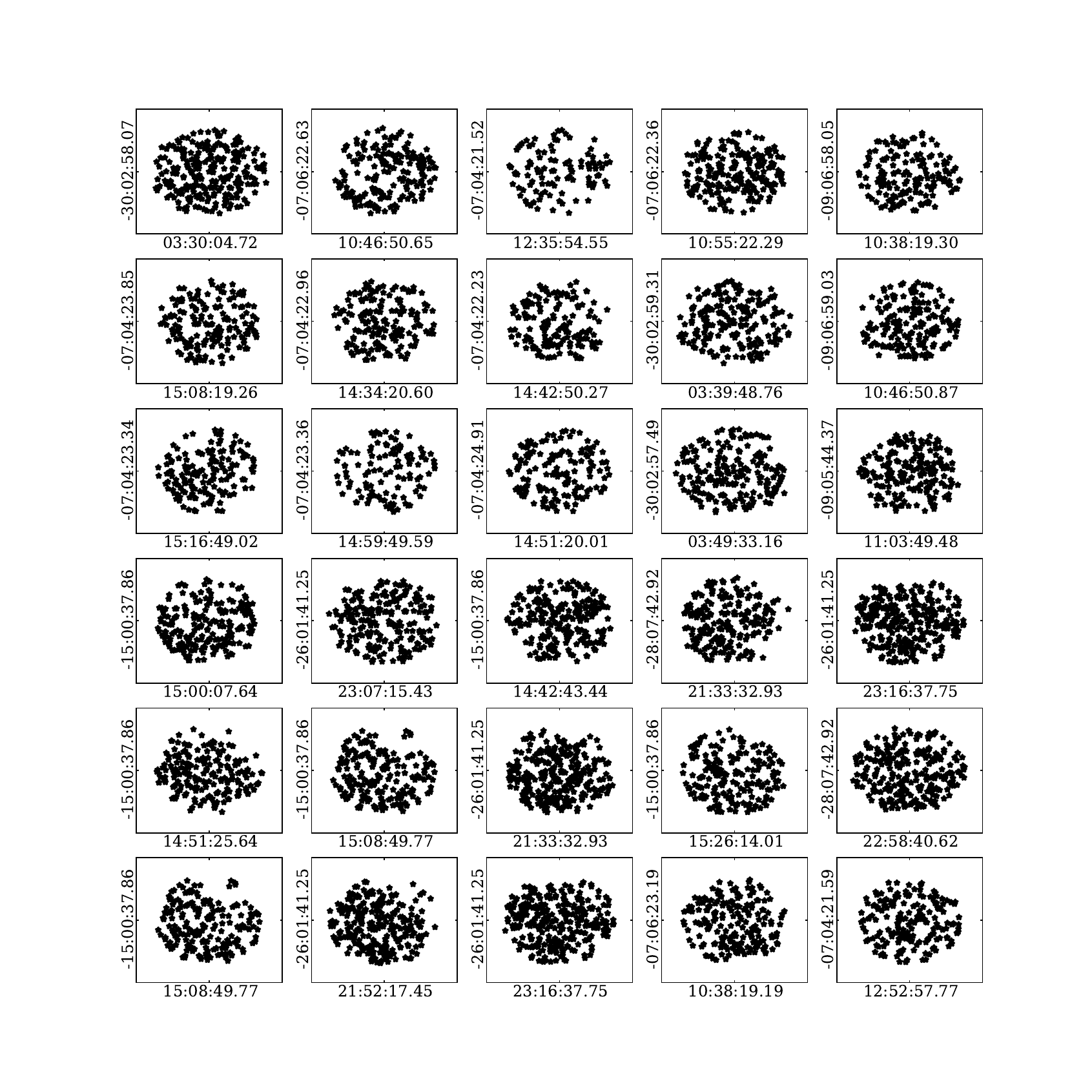}
\end{figure}
\twocolumn
\bsp	
\label{lastpage}
\end{document}